\documentclass[prb,twocolumn,showpacs,amsmath,amssymb,nofootinbib]{revtex4}
\usepackage{graphicx}
\usepackage{dcolumn}
\usepackage{bm}
\usepackage{color}
\usepackage{mathtools}
\usepackage{ulem}
\newcommand{\nn}{\nonumber}
\newcommand{\beq}{\begin{eqnarray}}
\newcommand{\eeq}{\end{eqnarray}}

\usepackage{atbegshi}
  \AtBeginShipoutFirst{}
\usepackage[colorlinks=true,linkcolor=blue,citecolor=blue,anchorcolor=blue,urlcolor=blue,pdfborderstyle={},dvipdfm]{hyperref}
\makeatletter
\AtBeginDocument{\@ifpackageloaded{natbib}{\ifNAT@numbers\if@filesw\immediate\write\@auxout{\string\global\string\NAT@numberstrue}\fi\fi}{}}
\makeatother

\begin{document}

\title{
Odd-frequency pairing and Ising spin susceptibility in time-reversal invariant superfluids and superconductors}

\author{Takeshi Mizushima}
\email{mizushima@mp.okayama-u.ac.jp}
\affiliation{Department of Physics, Okayama University, Okayama 700-8530, Japan} 
\date{\today}

\begin{abstract}
We here examine the relation between odd-frequency spin-triple even-parity (OTE) Cooper pairs and anomalous surface magnetic response in time-reversal invariant (TRI) spin-triplet superfluids and superconductors. The spin susceptibility generally consists of two contributions: Even-frequency odd-parity pair amplitudes and odd-frequency even-parity pair amplitudes. The OTE pair amplitudes are absent in the bulk region, but ubiquitously exist in the surface and interface region as Andreev bound states. We here clarify that additional discrete symmetries, originating from the internal symmetry and point group symmetry, impose strong constraint on the OTE pair amplitudes emergent in the surface of TRI superfluids and superconductors. As a result of the symmetry constraint, the magnetic response of the OTE pairs yields Ising-like anisotropy. For the topological phase of the $^3$He-B in a restricted geometry, the coupling of the OTE pair amplitudes to an applied field is prohibited by an additional discrete symmetry. Once the discrete symmetry is broken, however, the OTE pairs start to couple to the applied field, which anomalously enhances surface spin susceptibility. Furthermore, we extend this theory to TRI superconductors, where the corresponding discrete symmetry is the mirror reflection symmetry.
\end{abstract}


%67.30.H- Superfluid phase of 3He
%74.20.Rp Pairing symmetries (other than s-wave)
%74.20.-z Theories and models of superconducting state
%74.25.Dw Superconductivity phase diagrams
%74.45.+c Proximity effects; Andreev reflection; SN and SNS junctions
%74.81.-g Inhomogeneous superconductors and superconducting systems, including electronic inhomogeneities
%74.70.Ad Metals; alloys and binary compounds (including A15, MgB2, etc.)

\maketitle

%---------- Introduction
\section{Introduction}
Andreev bound states ubiquitously appear in inhomogeneous superconductors and superfluids including interfaces, surfaces, and vortices. The phase sensitive property has been utilized as a probe for the pair symmetry of host superconductors.~\cite{tanakaJPSJ2012,kashiwayaRPP2000} In addition, it has recently been unveiled that Andreev bound states have multi-faceted properties as odd-frequency Cooper pair correlation and Majorana fermions, giving rise to a drastic change of fundamental physical phenomena. 

Anomalous charge and spin transport, electromagnetic responses, proximity effects via Andreev bound states have  been clarified in light of odd-frequency Cooper pairing.~\cite{tanakaPRB1996,tanakaPRB1997,barashPRL1996,kashiwayaPRB1999,tanakaPRL2007,higashitaniJPSJ1997,WalterPRL1998,yokoyamaPRL2011,asanoPRL2011,bergeretPRL2001,tanakaPRB2005,tanakaPRL2007v2,linderPRL2009,linderPRB2010,higashitani2014,asano2014} In accordance with the Fermi-Dirac statistics, a wave function of Cooper pairs must change its sign after a permutation of two paired fermions. Then, as shown in Table~\ref{table1}, the pairing symmetry in a single-band superconductor is categorized to the four-fold way when the inversion symmetry is preserved. Two of them are even-frequency spin-singlet even-parity (ESE) and even-frequency spin-triplet odd-parity (ETO) pairings, which do not change the sign of Cooper pair wave function by the exchange of times of paired fermions. There still remain two possibilities of Cooper pair symmetries, odd-frequency spin-singlet odd-parity (OSO) and spin-triplet even-parity (OTE) pairs. Although conclusive evidence of odd-frequency pairing in bulk materials has not been observed experimentally since the first prediction by Berezinskii,~\cite{berezinskiiJETP1974} OSO and OTE pair amplitudes emerge ubiquitously in spatially non-uniform systems through Andreev bound states and anomalous proximity effect. 

Majorana fermions that are fermions equivalent to their own anti-particles are regarded as a special kind of Andreev bound states peculiar to topological superconductors and superfluids. One of the most distinctive characteristics of Majorana fermions is the Ising anisotropic magnetic response, {\it Majorana Ising spins}.~\cite{chungPRL2009,nagatoJPSJ2009,shindouPRB2010, volovikJETP2010, mizushimaPRL2012, tsutsumiJPSJ2013,shiozaki2014,mizushimaJPCM2014} Majorana Ising spins appear in the surface of time-reversal invariant (TRI) topological superconductors and superfluids, as a consequence of the chiral symmetry,~\cite{mizushimaPRL2012,shiozaki2014,mizushimaJPCM2014}  
\beq
\{\underline{\mathcal{H}}({\bm k}_{\perp}), \underline{\Gamma}\} = 0,
\label{eq:chiral}
\eeq 
where the chiral operator $\underline{\Gamma}$ that is anti-commutable with the Bogoliubov-de Gennes Hamiltonian $\underline{\mathcal{H}}({\bm k})$ obeys $\underline{\Gamma}^2=+1$. 
%The discrete symmetry has recently been a key concept to understanding topological superfluidity and superconductivity. 
The chiral operator $\underline{\Gamma}$ in Eq.~\eqref{eq:chiral} is constructed as a combination of the particle-hole symmetry and time-reversal symmetry for TRI superconductors and superfluids. The chiral symmetry can be constructed by combining the operators of the particle-hole symmetry ($\underline{\mathcal{C}}$), time-reversal symmetry ($\underline{\mathcal{T}}$), and a discrete symmetry ($\underline{\mathcal{M}}$), $\underline{\Gamma} = \underline{\mathcal{C}}\underline{\mathcal{T}}\underline{\mathcal{M}}$. The chiral symmetry may be preserved even if each discrete symmetry is broken.~\cite{mizushimaPRL2012,shiozaki2014,mizushimaJPCM2014}

A promising platform for realizing Majorana Ising spins is the superfluid $^3$He-B confined in a slab in which the chiral symmetry is preserved by a hidden order-two discrete symmetry even in the presence of a magnetic field.~\cite{mizushimaPRL2012,mizushimaJPCM2014} The other possible candidates have been proposed in topological crystalline superconductors including the E$_{1u}$ scenario of the heavy-fermion superconductor UPt$_3$,~\cite{tsutsumiJPSJ2013} a superconducting nanowire,~\cite{mizushimaNJP2013} noncentrosymmetric superconductors,~\cite{satoPRB2009} and the superconducting topological insulator Cu$_x$Bi$_2$Se$_3$.~\cite{fuPRL2010,sasakiPRL2011} In particular, although conflicting experimental results have been reported in Cu$_x$Bi$_2$Se$_3$,~\cite{sasakiPRL2011,kirzhner1,kirzhner2,kirzhner3,peng,levy} a consistent understanding is given in Ref.~\onlinecite{mizushima2014} in the context of the topological odd-parity superconductivity with the Fermi surface evolution. The discrete symmetry $\mathcal{M}$ originates from an internal symmetry in the case of $^3$He-B~\cite{mizushimaPRL2012,mizushimaJPCM2014} and a crystalline symmetry in superconductors.~\cite{mizushimaNJP2013,tsutsumiJPSJ2013,shiozaki2014} 

It has been widely accepted that the multi-faceted pictures of Andreev bound states as Majorana fermions and odd-frequency pairs are indispensable to understanding their anomalous contributions to magnetic responses, proximity effect, transport, and so on. In addition, a strong relation between the zero energy Majorana mode and odd-frequency pairing has been uncovered.~\cite{dainoPRB2012,asanoPRB2013,higashitaniPRB2012,tsutsumiJPSJ2012,stanev,hui} It is now important to establish unified understanding on Majorana Ising spins and the appearance of odd-frequency pairs in topological superfluids and superconductors. 

In this paper, we clarify the direct relation between odd-frequency Cooper pairs and Majorana Ising spins in TRI spin-triplet superfluids and superconductors. First, we derive the generic formalism for spin susceptibilities in TRI superconductors and superfluids in the frame of the quasiclassical theory. The spin susceptibility is separated to the contributions of ETO pairs and OTE pairs in the case of spin-triplet superconductors and superfluids. We here emphasize that a discrete symmetry originating from internal symmetry and crystalline symmetry imposes strong constraint on the Cooper pair symmetry emergent in a specular surface of TRI superfluids and superconductors. As a result of the symmetry constraint, the magnetic response of the OTE pairs yields Ising-like anisotropy. We also illustrate that once the OTE pairs are coupled to an applied field, they are responsible for anomalous enhancement of surface spin susceptibility, contrary to the contribution of ETO pairs. In this paper, special focus is placed on the surface spin susceptibility of the superfluid $^3$He-B and superconductor UPt$_3$ as concrete examples. 
 
%It has been unveiled that the Majorana Ising spin is in general a direct consequence of the chiral symmetry with $\Gamma^2=+1$ and a non-trivial winding number~\cite{mizushimaPRL2012} and the Ising spins are absent in superconductors with spontaneous breaking of the time-reversal symmetry. %In addition, topological phases protected by an internal or mirror symmetry may undergo a topological phase transition without closing the bulk excitation gap. In particular, in the case of $^3$He-B, the spontaneous breaking of a discrete internal symmetry triggers the topological phase transition. 

%It has recently been demonstrated that odd-frequency pairs are responsible for anomalous electromagnetic responses.~\cite{higashitani} 

%We here demonstrate that the former describes the rotation of the ${\bm d}$-vectors, while the contribution of odd-frequency even-parity Cooper pairs is associated with topological aspect of $^3$He-B under a magnetic field. Although the odd-frequency pairs always exist in the surface of $^3$He-B regardless of the applied field, they do not contribute to spin susceptibility in topological phase protected by pseudo-TRS. In non-topological phase with a ${\bm Z}_2$ symmetry breaking, however, the odd-frequency pairs are responsible for the strong enhancement of spin susceptibilities at the surface. 

This paper is arranged as follows: In Sec.~\ref{sec:spin}, we summarize the discrete symmetries in the context of the quasiclassical theory. We also derive the generic formalism of spin susceptibility, which is applicable not only to superfluids but also to the surface region of superconductors in the type-II limit. For TRI superfluids and superconductors, the spin susceptibility consists of the contributions from ETO and OTE pair amplitudes. In Sec.~\ref{sec:3He-B}, we examine the role of a hidden discrete symmetry on the odd-frequency pairs and the anomalous anisotropy of surface spin susceptibility in the superfluid $^3$He-B. In Sec.~\ref{sec:sc}, the theory is extended to TRI spin-triplet superconductors preserving a mirror reflection symmetry, such as UPt$_3$. The final section is devoted to conclusion and discussion. The details on the derivation of the gap equation in the presence of the dipole-dipole interaction are described in Appendix. Throughout this paper, we set $\hbar \!=\! k_{\rm B} \!=\! 1$ and the Pauli matrices $\underline{\tau}_{\mu}$ in particle-hole space and $\sigma _{\mu}$ in spin space. The repeated Greek indices imply the sum over $x$, $y$, and $z$. 

\begin{table}
\centering
\begin{tabular}{c|ccc|cc}
\hline\hline
& \multicolumn{3}{c}{Parity} 
& \multicolumn{2}{c}{Broken symmetry} \\
$\Delta$ & frequency & spin & parity & translational & time-inversion \\
\hline
ESE & $+$ & $-$ & $+$ & OSO & OTE \\
ETO & $+$ & $+$ & $-$ & OTE & OSO \\
OSO & $-$ & $-$ & $-$ & ESE & ETO \\
OTE & $-$ & $+$ & $+$ & ETO & ESE \\
\hline\hline
\end{tabular}
\caption{Classification of possible Cooper pairing in bulk superconductors: ESE, ETO, OSO, and OTE pairs. The fifth and sixth columns show the Cooper pair amplitudes emergent in systems with the breaking of translational symmetry and time-reversal symmetry, respectively.  
}
\label{table1}
\end{table}

%-------------------- 
\section{Spin susceptibility and Odd-frequency pairing}
\label{sec:spin}

The zero-energy density of states and Cooper pairing induced by surface and vortices can be directly associated with discrete symmetries, such as the time-reversal symmetry, particle-hole symmetry, and crystalline symmetries. In Sec.~\ref{sec:discrete}, we summarize the discrete symmetries of the Bogoliubov-de Gennes (BdG) Hamiltonian $\mathcal{H}({\bm k})$ and the quasiclassical propagator $\underline{g}$. Subsequently, we clarify that such discrete symmetry imposes strong constraint on odd-frequency pairing and gives rise to anomalous magnetic response. In Secs.~\ref{sec:3He-B} and \ref{sec:sc}, we will illustrate the remarkable consequence of the odd-frequency pair amplitudes, that is the Ising spin susceptibility in TRI superconductors and superfluids. 

\subsection{Discrete symmetries in quasiclassical theory}
\label{sec:discrete}

Before going into the symmetries in the quasiclassical formalism, let us start with the brief review on discrete symmetries of a general BdG Hamiltonian, which play a crucial role on determining the topologically nontrivial properties of superconductors and superfluids. The BdG Hamiltonian in bulk superconductors is in general given as 
\beq
\underline{\mathcal{H}}({\bm k}) = \left(
\begin{array}{cc}
\varepsilon({\bm k}) & \Delta ({\bm k}) \\ \Delta^{\dag}({\bm k}) & -\varepsilon^{\rm T}(-{\bm k})
\end{array}
\right), 
\label{eq:bdg}
\eeq
where $\varepsilon({\bm k})$ and $\Delta ({\bm k})$ are $2\times 2$ matrices in the spin space and we suppose $\Delta^{\rm T}(-{\bm k})=-\Delta({\bm k})$. %Note that the BdG Hamiltonian is related to the Green's function $\underline{G}$ as $\underline{G}^{-1} = i\omega _n - \underline{\mathcal{H}}$.

It is seen that $\underline{\mathcal{H}}({\bm k})$ in Eq.~\eqref{eq:bdg} has the particle-hole symmetry 
\beq
\underline{\mathcal{C}}~\underline{\mathcal{H}}({\bm k})\underline{\mathcal{C}}^{-1} 
= -\underline{\mathcal{H}}(-{\bm k}),
\label{eq:phs}
\eeq
where $\underline{\mathcal{C}} = \underline{\tau}_x K$ with $K$ being the complex conjugation operator converts the particle component of the quasiparticle wavefunction to the hole component and vice versa. 
In this paper, we consider TRI superfluids and superconductors, which yield $\Theta \Delta ({\bm k}) \Theta^{\rm T} = \Delta (-{\bm k})$. The time-reversal operator $\Theta$ is a unitary matrix and $\Theta^2 = -1$. When we also suppose the absence of time-reversal breaking perturbation, $\Theta \varepsilon ({\bm k}) \Theta^{\dag} = \varepsilon^{\ast}(-{\bm k})$, the BdG Hamiltonian $\underline{\mathcal{H}}({\bm k})$ preserves the time-reversal symmetry, 
\beq
\underline{\mathcal{T}}~\underline{\mathcal{H}}({\bm k})\underline{\mathcal{T}}^{-1} 
= \underline{\mathcal{H}}(-{\bm k}), \hspace{3mm}
\underline{\mathcal{T}} = 
\left( 
\begin{array}{cc}
\Theta & 0 \\ 0 & \Theta^{\ast} 
\end{array}
\right)K.
\eeq

In addition to such fundamental symmetry, the pair potential may hold the discrete rotational symmetry in spin and momentum spaces.~\cite{mizushimaPRL2012} Specifically, the B-phase of bulk superfluid $^3$He is invariant under the joint rotation of spin and orbital spaces ${\rm SO}(3)_{{\bm L}+{\bm S}}$. The $\pi$-rotation is then defined as a subgroup of the continuous symmetry group. In this case, we obtain $U(\pi)\Delta ({\bm k}) U^{\rm T}(\pi) = \Delta (R{\bm k})$ with the $\pi$-rotation operators $U(\pi)$ in spin space and $R$ in momentum space. For the case that $U(\pi)\varepsilon({\bm k})U^{\dag}(\pi) = \varepsilon (R{\bm k})$, the BdG Hamiltonian satisfies the $\pi$-rotational symmetry,~\cite{mizushimaPRL2012,mizushimaJPCM2014}
\beq
\mathcal{U}(\pi) \mathcal{H}({\bm k})\mathcal{U}^{\dag}(\pi) = \mathcal{H}(R{\bm k}),
\label{eq:pi}
\eeq
where $\underline{\mathcal{U}}(\pi) = {\rm diag}[U(\pi), U^{\ast}(\pi)]$. 

It is also found that superconducting states retain the mirror symmetry if the pair potential is odd or even under the mirror reflection, $M\Delta ({\bm k})M^{\rm T} = \pm \Delta (\underline{\bm k}_{\rm M})$.~\cite{uenoPRL2013} Here, we define the mirror operator, $M = i({\bm \sigma}\cdot\hat{\bm o})$, where $\hat{\bm o}$ is the unit vector normal to the mirror plane. The mirror operator changes the spin ${\bm \sigma}\rightarrow -{\bm \sigma}+2\hat{\bm o}({\bm \sigma}\cdot\hat{\bm o})$ and the momentum ${\bm k}\rightarrow\underline{\bm k}_{\rm M} = {\bm k}-2\hat{\bm o}({\bm k}\cdot\hat{\bm o})$. When the normal state has the mirror symmetry, $M\varepsilon ({\bm k})M^{\dag} = \varepsilon (\underline{\bm k}_{\rm M})$, the BdG Hamiltonian holds the mirror reflection symmetry, 
\beq
\underline{\mathcal{M}}^{\eta}\mathcal{H}({\bm k})\underline{\mathcal{M}}^{\eta \dag} 
= \underline{\mathcal{H}}(\underline{\bm k}_{\rm M}),
\label{eq:Hmirror}
\eeq
where
\beq
\underline{\mathcal{M}}^{\eta} = 
\left( 
\begin{array}{cc}
M & 0 \\ 0 & \eta M^{\ast}
\end{array}
\right), \hspace{3mm} \eta = \pm
\label{eq:mirror}
\eeq
Note that even if each discrete symmetry is broken, the BdG Hamiltonian $\underline{\mathcal{H}}({\bm k})$ may still preserve a discrete symmetry constructed by combining some of discrete symmetries.

The equilibrium properties of superfluids and superconductors are well describable with the quasiclassical theory,~\cite{serene} which is reliable in $T_{\rm c0}\ll T_{\rm F}$ ($T_{\rm c0}$ is the transition temperature of the bulk $^3$He-B). The central object is the quasiclassical Green's functions $\underline{g} \!\equiv\! \underline{g}(\hat{\bm k},{\bm r};\omega _n)$, which are obtained by integrating the Matsubara Green's function $\underline{G}$ over a shell $v_{\rm F}|k-k_{\rm F}| < E_{\rm c} \ll E_{\rm F}$, 
\beq
\underline{g}(\hat{\bm k},{\bm r};\omega _n) = \frac{1}{a} \int^{+E_{\rm c}}_{-E_{\rm c}} d\xi _{\bm k}
\underline{\tau}_z\underline{G}({\bm k},{\bm r};\omega _n).
\eeq
The normalization constant $a$ corresponds to the weight of the quasiparticle pole in the spectral function and the Matsubara frequency is $\omega _n \!=\! (2n+1)\pi T$ with $n\!\in\! \mathbb{Z}$. The full propagator is defined with the Nambu spinor of the fermionic field operators ${\bm \Psi}=(\psi _{\uparrow},\psi _{\downarrow},\psi^{\dag}_{\uparrow},\psi^{\dag}_{\downarrow})^{\rm T}$ by
\begin{align}
\underline{G}({\bm k},{\bm r};\omega _n) 
=& -\int^{\beta}_0 d\tau e^{i\omega _n \tau}
\int d{\bm r}_{12} e^{-i{\bm k}\cdot{\bm r}_{12}} \nn \\
& \times \left\langle
{\rm T}_{\tau}{\bm \Psi}\left({\bm r}_+,\tau\right)\bar{\bm \Psi}\left({\bm r}_-,0\right)
\right\rangle
\label{eq:originalG}
\end{align}
with ${\bm r}_{\pm} \equiv {\bm r}\pm {\bm r}_{12}/2$ and $\beta^{-1} \equiv T$. The quasiclassical propagator $\underline{g}$ that is a $4\times 4$ matrix is parameterized with Pauli matrices in spin space $\sigma _{\mu}$ as
\beq
\underline{g} = \left(
\begin{array}{cc}
g_{0} + {\sigma}_{\mu} g_{\mu} & i\sigma _y f_0 + i {\sigma}_{\mu} {\sigma}_y f_{\mu}  \\ 
i\sigma _y \bar{f}_0 +i \sigma _y {\sigma}_{\mu}\bar{f}_{\mu}  & \bar{g}_{0} + {\sigma}^{\rm T}_{\mu} \bar{g}_{\mu}
\end{array}
\right).
\label{eq:g}
\eeq
Here, $\sigma^{\rm T}_{\mu}$ denotes the transpose of the Pauli matrices $\sigma _{\mu}$. The off-diagonal propagators are composed of spin-singlet and triplet Cooper pair amplitudes, $f_0$ and $f_{\mu}$. 

The quasiclassical propagator $\underline{g}\equiv\underline{g}(\hat{\bm k},{\bm r};\omega _n)$ is governed by the Eilenberger equation,~\cite{serene} 
\beq
[i\omega _n \underline{\tau}_z - \underline{v}(\hat{\bm k},{\bm r})
- \underline{\Delta}(\hat{\bm k},{\bm r}), 
\underline{g}
] \!=\!- i {\bm v}_{\rm F} \!\cdot{\bm \nabla}\!
\underline{g}.
\label{eq:eilen}
\eeq
with the normalization condition, 
\beq
\left[ \underline{g} (\hat{\bm k},{\bm r};\omega _n)\right]^2 \!=\! -\pi^2,
\label{eq:norm}
\eeq 
where we introduce 
\beq
\underline{\Delta}(\hat{\bm k},{\bm r}) 
= \left(
\begin{array}{cc}
& \Delta(\hat{\bm k},{\bm r}) \\
\Delta^{\dag}(-\hat{\bm k},{\bm r}) & 
\end{array}
\right). 
\eeq
The term $ \underline{v}$ in Eq.~(\ref{eq:eilen}) consists of an external potential $\underline{v}_{\rm ext}$ and quasiclassical self-energy associated with Fermi liquid corrections $\underline{\nu}$, as $\underline{v}(\hat{\bm k},{\bm r}) = \underline{v}_{\rm ext}({\bm r}) + \underline{\nu}(\hat{\bm k},{\bm r})$, where 
\beq
\underline{\nu} = 
\left(
\begin{array}{cc}
\nu _{0} + {\sigma}_{\mu} \nu _{\mu} &   \\ 
 & \bar{\nu}_{0} + {\sigma}^{\rm T}_{\mu} \bar{\nu}_{\mu}
\end{array}
\right)
\eeq
The quasiclassical propagators also satisfy the following relations arising from the Fermi statistics in Eq.~\eqref{eq:originalG},
\begin{gather}
\left[ \underline{g}(\hat{\bm k},{\bm r};\omega_n)\right]^{\dag} 
= \underline{\tau}_z\underline{g}(\hat{\bm k},{\bm r};-\omega_n) \underline{\tau}_z, 
\label{eq:sym1} \\
\left[ \underline{g}(\hat{\bm k},{\bm r};\omega_n)\right]^{\rm T} 
= \underline{\tau}_y\underline{g}(-\hat{\bm k},{\bm r};-\omega_n) \underline{\tau}_y.
\label{eq:sym2}
\end{gather}
It is important to mention that the normalization condition, $gf = - f\bar{g}$ and $\bar{g}\bar{f} = - \bar{f}g$, leads to the relation,
\beq
\bar{g}_0(\hat{\bm k},{\bm r};\omega _n) = -g_0(\hat{\bm k},{\bm r};\omega _n).
\label{eq:trs2}
\eeq

The discrete symmetries which are preserved by the BdG Hamiltonian are extended to the quasiclassical formalism, which add constraint on the quasiclassical propagator.
First, the particle-hole symmetry in Eq.~\eqref{eq:phs} is recast into 
\beq
\underline{\mathcal{C}}~\underline{g}(\hat{\bm k},{\bm r};\omega _n) \underline{\mathcal{C}}^{-1}
= \underline{g}(-\hat{\bm k},{\bm r};\omega _n).
\eeq
This symmetry can be obtained from the basic relations of the quasiclassical propagator in Eqs.~\eqref{eq:sym1} and \eqref{eq:sym2}. For time-reversal invariant superconductors and superfluids which yeild $\Theta \Delta ({\bm k}) \Theta^{\rm T} = \Delta (-{\bm k})$, the time-reversal symmetry is 
\beq
\underline{\mathcal{T}}~
\underline{g}(\hat{\bm k},{\bm r};\omega _n) \underline{\mathcal{T}}^{-1}
= \underline{g}(-\hat{\bm k},{\bm r};-\omega _n) , 
\label{eq:trs1}
\eeq
where we also suppose that $\underline{v}$ does not contains the time-reversal breaking term. Similarly, the $\pi$-rotational symmetry in Eq.~\eqref{eq:pi} and mirror symmetry in Eq.~\eqref{eq:mirror} are recast into 
\beq
\underline{\mathcal{U}}(\pi)\underline{g}(\hat{\bm k},{\bm r};\omega _n) \underline{\mathcal{U}}^{\dag}(\pi)
= \underline{g}(R\hat{\bm k},R{\bm r};\omega _n),
\label{eq:pi2}
\eeq
and 
\beq
\underline{\mathcal{M}}^{\eta}\underline{g}(\hat{\bm k},{\bm r};\omega _n) 
\underline{\mathcal{M}}^{\eta\dag}
= \underline{g}(\underline{\hat{\bm k}}_{\rm M},\underline{\bm r}_{\rm M};\omega _n).
\label{eq:gmirror}
\eeq
We will show in Secs.~\ref{sec:3He-B} and \ref{sec:sc} that these discrete rotational symmetries add a strong constraint to Cooper pairings induced in a specular surface and give rise to the Ising anisotropy of surface spin susceptibility. 

\subsection{Spin susceptibility in quasiclassical thory}

We here derive the generic form of the magnetization density $M_{\mu}({\bm r})$ for superfluids under a spatially uniform magnetic field ${\bm H}\!=\!H\hat{\bm h}$. In this situation, the potential term $\underline{v}(\hat{\bm k},{\bm r})$ in the quasiclassical equation \eqref{eq:eilen} is composed of a magnetic Zeeman field and quasiclassical self-energies $\underline{\nu}$,
\beq
\underline{v}(\hat{\bm k},{\bm r}) = - \frac{1}{1+F^{\rm a}_0}\mu _{\rm n} H_{\mu}\left(
\begin{array}{cc}
\sigma _{\mu} & \\ &  \sigma^{\rm T}_{\mu} 
\end{array}
\right) + \underline{\nu}(\hat{\bm k},{\bm r}), 
\label{eq:v2}
\eeq
where $F^{\rm a}_0$ is the Fermi liquid parameter associated with the enhancement of spin susceptibility and $\mu _{\rm n}$ is the magnetic moment of $^3$He nuclei.~\cite{vollhardt} In the quasiclassical formalism, the magnetization density is given by~\cite{serene,mizushimaPRL2012,mizushimaPRB2012,mizushimaJPCM2014}
\beq
M_{\mu}(z) = M_{\rm N}\left[
\hat{h}_{\mu} + \frac{1}{\mu _{\rm n} H}\langle g_{\mu}(\hat{\bm k},z;\omega_n)\rangle _{\hat{\bm k},n}
\right].
\label{eq:M}
\eeq
This is also applicable to the surface region of superconductors in the type-II limit where the surface region within the coherence length $\xi$ is much thinner than the penetration depth of the external field. For superconductors, $\mu _{\rm n}$ in Eqs.~\eqref{eq:v2} and \eqref{eq:M} is replaced to the Bohr magneton $\mu _{\rm B}$ and the applied field ${\bm H}$ is replaced to the internal field ${\bm B}$. In Eq.~(\ref{eq:M}), we introduce the average over the Fermi surface
$\langle\cdots\rangle _{\hat{\bm k},n} = \frac{T}{N_{\rm F}}\sum _{n}
\int \frac{d\hat{\bm k}}{(2\pi)^3 |{\bm v}_{\rm F}(\hat{\bm k})|}\cdots$ ,
where $N_{\rm F} \!=\! \int \frac{d\hat{\bm k}}{(2\pi)^3|{\bm v}_{\rm F}(\hat{\bm k})|}$ is the total density of states at the Fermi surface in the normal state and the Fermi velocity at ${\bm k} \!=\! k_{\rm F}\hat{\bm k}$ is defined as ${\bm v}_{\rm F}(\hat{\bm k}) \!=\! \partial \epsilon ({\bm k})/\partial {\bm k}|_{{\bm k}=k_{\rm F}\hat{\bm k}}$. For three dimensional Fermi sphere, one finds ${\bm v}_{\rm F}(\hat{\bm k}) \!=\! v_{\rm F}\hat{\bm k}$ and $N_{\rm F} \!=\! \frac{1}{2\pi^2 v_{\rm F}}$, which reduces to $\langle\cdots\rangle _{\hat{\bm k},n} = T\sum _{n} \int \frac{d{\bm k}}{4\pi}\cdots $. 
The magnetization in normal $^3$He is
$M_{\rm N} \!=\! \chi _{\rm N}H \!=\! \frac{2\mu^2_{\rm n}}{1+F^{\rm a}_0}N_{\rm F}H$. 

The quasiclassical propagator must satisfy a constraint given in Eq.~(\ref{eq:norm}) which requires the propagators to hold the relation, 
$g_{\mu} \!=\! ( f_0 \bar{f}_{\mu} + \bar{f}_0 f_{\mu}
+ i\epsilon _{\mu\nu\eta}f_{\nu}\bar{f}_{\eta})/2g_0$. This relates the spin component of quasiclassical propagators to spin-singlet and -triplet Cooper pair amplitudes. %Using the relation, the magnetization density reduces to
%\beq
%\frac{M_{\mu}({\bm r})}{M_{\rm N}} = \hat{h}_{\mu} + \frac{1}{\mu _{\rm n}H}
%\left\langle \frac{f_0 \bar{f}_{\mu} + \bar{f}_0 f_{\mu}+ i\epsilon _{\mu\nu\eta}f_{\nu}\bar{f}_{\eta}}{2g_0} \right\rangle _{\hat{\bm k},n}.
%\label{eq:m}
%\eeq
Using the relation and the symmetries in Eqs.~(\ref{eq:sym1}), (\ref{eq:sym2}), and (\ref{eq:trs2}), the magnetization density in Eq.~(\ref{eq:M}) reduces to
\beq
\frac{M_{\mu}({\bm r})}{M_{\rm N}} = \hat{h}_{\mu} + \frac{1}{\mu _{\rm n}H}
\left\langle \frac{f_0 \bar{f}_{\mu} + \bar{f}_0 f_{\mu}}{2g_0} \right\rangle _{\hat{\bm k},n}.
\label{eq:m2}
\eeq
This indicates that only the mixing term of spin-singlet and triplet Cooper pair amplitudes contributes to the spin susceptibilities. This expression is a quite generic form for $M_{\mu}({\bm r})$ in superfluids and also applicable to the surface region of type-II superconductors. This was first derived in Ref.~\onlinecite{higashitaniPRL2013} for the aerogel-superfluid $^3$He-B system.

%-------------------- 
\subsection{Odd-frequency pairs and spin susceptibility}
\label{sec:susceptibility}

In general, the Cooper pair amplitudes are separated to even-frequency and odd-frequency components, $
f_{\mu} \!=\! f^{{\rm EF}}_{\mu} + f^{{\rm OF}}_{\mu}$ and $f_{0} \!=\! f^{{\rm EF}}_{0} + f^{{\rm OF}}_{0}$, where even- and odd-frequency pair amplitudes are defined as ($j \!=\! 0, x, y, z$)
\begin{gather}
f^{\rm EF}_{j}(\hat{\bm k},{\bm r};\omega _n) = \frac{1}{2} \left[
f_{j}(\hat{\bm k},{\bm r};\omega _n) + f_{j}(\hat{\bm k},{\bm r};-\omega _n)
\right], 
\label{eq:even} \\
f^{\rm OF}_{j}(\hat{\bm k},{\bm r};\omega _n) = \frac{1}{2} \left[
f_{j}(\hat{\bm k},{\bm r};\omega _n) - f_{j}(\hat{\bm k},{\bm r};-\omega _n)
\right].
\label{eq:odd}
\end{gather}
In the case of spin-triplet superconductors and superfluids, ETO components $f^{{\rm EF}}_{\mu}$ exist in the bulk and an applied magnetic field induces OSO pairs $f^{{\rm OF}}_{0}$. The other components, $f^{\rm EF}_0$ and $f^{\rm OF}_{\mu}$, do not play an important role in the bulk of spin-triplet superfluids and superconductors. As summarized in Table~\ref{table1}, however, the translational symmetry breaking due to a surface boundary condition and vortices, induces OTE components $f^{\rm OF}_{\mu}$ even in the zero field limit.~\cite{tanakaJPSJ2012,tanakaPRL2007,tanakaPRB2007,yokoyamaPRB2008,yokoyamaJPSJ2010,dainoPRB2012} Note that the OTE Cooper pair amplitudes $f^{\rm OF}_{\mu}$ are associated with the low-energy density of states originating from the surface Andreev bound states.~\cite{higashitaniPRB2012,tsutsumiJPSJ2012} Furthermore, at the zero energy limit, $f^{\rm OF}_{\mu}$ is equivalent to the Majorana zero modes.~\cite{dainoPRB2012,asanoPRB2013} 

We now clarify the relation between OTE Cooper pairs and spin susceptibility in spin-triplet superfluids and superconductors. We here deal with a magnetic Zeeman field perturbatively in parameter, $\mu _{\rm n}H/\Delta \!\ll\! 1$. Then, we formally expand $g_0$, $f_0$, and $f_{\mu}$ in powers of $\mu _{\rm n}H/\Delta$: $g_0 \!=\! g^{(0)}_0 + g^{(1)}_0 +\cdots$, $f_0 \!=\! f^{(1)}_0 + \cdots$, and $f_{\mu} \!=\! f^{(0)}_{\mu} + f^{(1)}_{\mu} + \cdots$. At zero fields, TRI superfluids and superconductors hold the time-reversal symmetry (\ref{eq:trs1}). 
Combining the symmetric property in Eq.~(\ref{eq:trs1}) with Eqs.~(\ref{eq:sym1}), (\ref{eq:sym2}), and (\ref{eq:trs2}), therefore, one finds 
\beq
g^{(0)}_0(\hat{\bm k},z; \omega _n) = - g^{(0)}_0(\hat{\bm k},z; -\omega _n).
\label{eq:g0} 
\eeq
Substituting all these in Eq.~(\ref{eq:m2}) and using the symmetry in Eq.~(\ref{eq:g0}), one finds that the spin susceptibility $\chi \equiv \hat{h}_{\mu}\chi _{\mu\nu}\hat{h}_{\nu}$ is composed of the contributions of odd- and even-parity Cooper pair amplitudes, 
\beq
\chi(z) 
= \chi _{\rm N} + \chi^{\rm OP}(z) + \chi^{\rm EP}(z).
\label{eq:chi2}
\eeq
The spin susceptibility tensor $\chi _{\mu\nu}$ is defined as $M_{\mu} = \chi _{\mu \nu}H_{\nu}$. The odd-parity contribution $\chi^{\rm OP}(z)$ is given by the mixing term of the OSO pair amplitude $f^{\rm OF}_0$ and the ETO pair ${\bm f}^{\rm EF}$,
\beq
\frac{\chi^{\rm OP}(z)}{\chi _{\rm N}} \equiv
\frac{1}{\mu _{\rm n}H}{\rm Re}\left\langle 
\frac{f^{{\rm OF}(1)}_0\hat{h}_{\mu}f^{{\rm EF}(0)\ast}_{\mu}}{g^{(0)}_0}
\right\rangle _{\hat{\bm k},n}. \label{eq:chiOP} 
\eeq
The even-parity contribution $\chi^{\rm EP}(z)$ is given by the mixing term of the ESE pair amplitude $f^{\rm EF}_0$ and the OTE pair ${\bm f}^{\rm OF}$,
\beq
\frac{\chi^{\rm EP}(z)}{\chi _{\rm N}} \equiv
-\frac{1}{\mu _{\rm n}H}{\rm Re}\left\langle 
\frac{f^{{\rm EF}(1)}_0\hat{h}_{\mu}f^{{\rm OF}(0)\ast}_{\mu}}{g^{(0)}_0}
\right\rangle _{\hat{\bm k},n}. \label{eq:chiEP}
\eeq
This indicates that the spin susceptibility is separated to the contributions from odd-parity Cooper pair amplitudes, $\chi^{\rm OP}$, and even-parity pairing, $\chi^{\rm EP}$, which mix the field-induced spin-singlet pairing $f^{(1)}_{0}$ and spin-triplet pairing at zero fields $f^{(0)}_{\mu}$. The spin-triplet pairings $f^{(0)}_{\mu}$ at zero fields are directly coupled to the applied field. Note that only the ETO pairings $f^{{\rm EF}(0)}_{\mu}$ remains finite in the bulk of spin-triplet superfluids and superconductors and the behavior of $\chi^{\rm OP}$ is understandable with the rotation of the ${\bm d}$-vector, which is responsible for the diamagnetic response $\chi^{\rm OP} \le 0$. In contrast, the OTE Cooper pairs $f^{{\rm OF}(0)}_{\mu}$ are absent in the bulk and induced by the breaking of translational symmetry at surfaces, interfaces, or vortices. Therefore, the spin susceptibility at surfaces is determined by the OTE pairing $f^{(0)}_{\mu}$ directly coupled to the applied field in addition to the ordinary contribution to the ${\bm d}$-vectors. It has recently been clarified that odd-frequency pairs increase the spin susceptibility in the case of spin singlet superconductors.~\cite{higashitani2014}

%-------------------- 
\section{Odd-frequency pairs and Majorana Ising spin in $^3$He-B}
\label{sec:3He-B}

In this section, we consider the spin susceptibility in the B-phase of spin-triplet superfluid $^3$He confined in a restricted geometry. The geometry is illustrated in Fig.~\ref{fig:slab}. As shown in Eqs.~\eqref{eq:chiOP} and \eqref{eq:chiEP}, the spin susceptibility in TRI superfluids is determined by the pair amplitudes in the absence of a magnetic field, $f^{(0)}_{\mu}$. In particular, we will show that the OTE pairing, $f^{(0)\rm OF}_{\mu}$ plays a key role. It is also clarified that the discrete symmetries add a strong constraint on the emergent pair amplitudes $f^{(0)}_{\mu}$ at surfaces, which gives rise to Ising anisotropy of surface magnetic response.

\subsection{Symmetry and odd-frequency pairings at zero fields}

We start by summarizing the remaining discrete symmetry in superfluid $^3$He-B confined in a slab geometry. The symmetry group relevant to the normal $^3$He in the bulk is given as $G={\rm SO}(3)_{\bm S}\times {\rm SO}(3)_{\bm L} \times {\rm U}(1)_{\phi} \times {\rm T} \times {\rm C}$, where ${\rm SO}(3)_{\bm S}$ and ${\rm SO}(3)_{\bm L}$ denote the three-dimensional rotational symmetry in spin and coordinate spaces, and ${\rm T}$ and ${\rm C}$ are the time-reversal and particle-hole symmetries, respectively. The bulk B-phase retains $H={\rm SO}(3)_{{\bm S}+{\bm L}}\times {\rm T}\times {\rm C}$, which is the maximal subgroup of $G$. The degeneracy space is characterized with the relative rotation ${\rm SO}(3)_{{\bm S}-{\bm L}}$ as $R=G/H={\rm SO}(3)_{{\bm S}-{\bm L}}\times {\rm U}(1)_{\phi}$. Then, the pair potential is expressed as $\Delta ({\bm k},{\bm r}) = i\sigma _{\mu} \sigma _y d_{\mu}({\bm k},{\bm r})$, where the ${\bm d}$-vector of the B-phase is given by 
\beq
d_{\mu}(\hat{\bm k},{\bm r}) = R_{\mu\nu}(\hat{\bm n},\varphi)d_{\mu \nu}({\bm r})\hat{k}_{\nu},
\label{eq:dvec}
\eeq
where $R_{\mu\nu} \!\in\! {\rm SO}(3)_{{\bm S}-{\bm L}}$ and we omit the ${\rm U}(1)$ phase for simplicity. The rotation axis $\hat{\bm n}$, the angle $\varphi$, and the order parameter amplitudes $\Delta _{\mu}$ are obtained as the self-consistent solution of the quasiclassical equation coupled with the gap equation and Fermi liquid corrections. The $\hat{\bm n}$-texture is supposed to be spatially uniform, which is forced by the confinement. The specular surfaces at $z_0 = 0$ and $D$ impose the quasiclassical propagators on the boundary condition, 
\beq
\underline{g}(\hat{\bm k},z = z_0; \omega _n) = \underline{g}(\underline{\hat{\bm k}},z = z_0; \omega _n),
\label{eq:bc}
\eeq
where $\underline{\hat{\bm k}}\equiv \hat{\bm k} - 2 \hat{\bm z} (\hat{\bm z}\cdot\hat{\bm k})$ is the momentum specularly reflected by the surface. 

%-------------------------------------------------------------
\begin{figure}[t!]
\includegraphics[width=60mm]{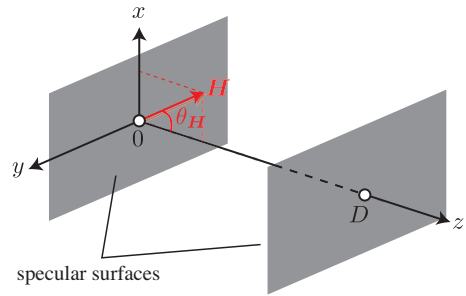}
\caption{(color online) Schematic picture of the geometry considered here. The liquid $^3$He is confined to a slab geometry within $z\in[0,D]$, where the two specular surfaces are situated at $z =z_0= 0$ and $D$. }
\label{fig:slab}
\end{figure}
%------------------------------------------------------------/

%We here clarify that pair amplitudes in the absence of a magnetic field, $f^{(0)}_{\mu}$, are determined by the discrete symmetry described in Eqs.~\eqref{eq:pi} and \eqref{eq:pi2}, which arises from the continuous rotational symmetry of $d_{\mu}(\hat{\bm k},{\bm r})$ in the slab geometry. 

The confinement in Fig.~\ref{fig:slab} reduces ${\rm SO}(3)_{\bm L}$ to ${\rm SO}(2)_{L_z}$ that is the rotational symmetry about the surface normal axis. The symmetry group in normal $^3$He confined in this geometry is $G_{\rm slab} = {\rm SO}(3)_{\bm S}\times {\rm SO}(2)_{L_z} \times {\rm U}(1)_{\phi} \times {\rm T} \times {\rm C}$. The pair potential relevant to this situation is given with $d_{\mu \nu}({\bm r})= R_{\mu\eta}(\hat{\bm n},\varphi)d_{\eta \nu}(z)$ in Eq.~\eqref{eq:dvec} as~\cite{vollhardt}
\beq
d_{\mu \nu}(z)  = \Delta _{\parallel}(z) \left( \delta _{\mu \nu} - \hat{z}_{\mu}\hat{z}_{\nu}
\right) + \Delta _{\perp}(z) \hat{z}_{\mu} \hat{z}_{\nu},
\label{eq:dvec_slab}
\eeq
where without loss of generality, we set $\Delta _{\parallel}\!\in\! \mathbb{R}$ and $\Delta _{\perp}\!\in\!\mathbb{R}$. In such a geometry, the B-phase is still invariant under the simultaneous rotation in spin and orbital spaces about the surface normal axis, $H_{\rm slab} = {\rm SO}(2)_{S_z+L_z}\times {\rm T}\times {\rm C}$, where the ${\rm SO}(2)_{{\bm S}+{\bm L}}$ symmetry is expressed as
\beq
U(\phi) \Delta(\hat{\bm k},z) U^{\rm T}(\phi)= \Delta(O\hat{\bm k},z),
\label{eq:pi3}
\eeq
where $(O)_{\mu\nu}\equiv R_{\mu\nu}(\hat{\bm z},\phi)$ denotes the rotational matrix about the $\hat{\bm z}$-axis by any angle $\phi$. The $2\times 2$ matrix in spin space, $U(\phi)$, is the corresponding rotational matrix in spin space and is defined as 
\beq
U(\phi) = U(\hat{\bm n},\varphi) e^{-i\phi\sigma _z/2} U^{\dag}(\hat{\bm n},\varphi),
\eeq
where $e^{-i\phi\sigma _z/2}$ is the ${\rm SU}(2)$ representation of $(O)_{\mu\nu}\equiv R_{\mu\nu}(\hat{\bm z},\phi)$.

The $\pi$-rotational symmetry $U(\pi)$ in Eq.~\eqref{eq:pi} is defined as the subgroup of ${\rm SO}(2)_{S_z+L_z}$, $U(\pi)\equiv U(\phi \!=\! \pi)$. This imposes the additional discrete symmetry on the quasiclassical propagator as shown in Eq.~\eqref{eq:pi2}, 
\beq
\underline{\mathcal{U}}(\pi)
\underline{g}^{(0)}(\hat{\bm k},z;\omega _n) \underline{\mathcal{U}}^{\dag}(\pi)
= \underline{g}^{(0)}(-\underline{\hat{\bm k}},z;\omega _n),
\label{eq:pi4}
\eeq
where $\underline{\mathcal{U}}(\pi) = {\rm diag}[U(\pi),U^{\ast}(\pi)]$. Combining this with the boundary condition in Eq.~\eqref{eq:bc} and the relation in Eq.~\eqref{eq:sym2}, one obtains the relation between $\underline{g}(\omega _n)$ and $\underline{g}(-\omega _n)$ at the surface $z=z_0$ as
\begin{align}
\underline{g}^{(0)}(\hat{\bm k},z_0;-\omega _n) = 
\underline{\mathcal{U}}(\pi)\underline{\tau}_y
\left[\underline{g}^{(0)}(\hat{\bm k},z_0;\omega _n)\right]^{\rm T} \underline{\tau}_y
\underline{\mathcal{U}}^{\dag} (\pi).
\label{eq:relation_g0}
\end{align}

It is convenient to introduce $\underline{\tilde{g}}^{(0)}$ obtained by the unitary transformation of the original quasiclassical propagator as
\beq
\underline{\tilde{g}}^{(0)}(\hat{\bm k},z;\omega _n) = 
\underline{\mathcal{U}}^{\dag}(\hat{\bm n},\varphi) 
\underline{g}^{(0)}(\hat{\bm k},z;\omega _n) \underline{\mathcal{U}}(\hat{\bm n},\varphi). 
\eeq
The propagator $\underline{\tilde{g}}^{(0)}$ obeys the quasiclassical equation with the definition $\underline{\tilde{\Delta}} \equiv U^{\dag}(\hat{\bm n},\varphi)\Delta (\hat{\bm k},z)U(\hat{\bm n},\varphi)^{\ast} = i\sigma _{\mu}\sigma _y d_{\mu \nu}(z)\hat{k}_{\nu}$, 
$
[ i\omega _n \underline{\tau}_z - \underline{\tilde{\nu}}(\hat{\bm k},z) - \underline{\tilde{\Delta}}(\hat{\bm k},z),
\underline{\tilde{g}}^{(0)} ] \!=\! - i{\bm v}_{\rm F}\cdot {\bm \nabla}\underline{\tilde{g}}^{(0)} 
$.
This is equivalent to Eq.~\eqref{eq:eilen} in the case of $\hat{\bm n} = \hat{\bm z}$ and $\varphi =0$. Then, Eq.~\eqref{eq:relation_g0} imposes the constraint on the pair amplitudes $\tilde{f}_{\mu}$ at the surfaces as
\beq
\tilde{f}^{{\rm OF}(0)}_{\parallel}(\theta _{\bm k},z_0;\omega _n) = \tilde{f}^{{\rm EF}(0)}_{z}(\theta _{\bm k},z_0;\omega _n) = 0,
\label{eq:tildef}
\eeq
where $(\tilde{f}^{(0)}_x,\tilde{f}^{(0)}_y) = \tilde{f}^{(0)}_{\parallel}(\cos\phi _{\bm k},\sin\phi _{\bm k})$ results from the ${\rm SO}(2)_{S_z+L_z}$ symmetry. It turns out from Eq.~\eqref{eq:tildef} that at the surface, only $\tilde{f}_{z}$ has odd-frequency Cooper pairs at the surfaces and the even-frequency Cooper pairs survive only in $\tilde{f}_{\parallel}$. By using this notation, the pair amplitudes $f^{(0)}_{\mu}\equiv f^{(0)}_{\mu}(\hat{\bm k},z_0;\omega _n)$ at the surfaces of superfluid $^3$He-B are expressed as 
\begin{gather}
f^{{\rm EF}(0)}_{\mu} = \left( R_{\mu x}(\hat{\bm n},\varphi)\cos\phi _{\bm k} 
+ R_{\mu y}(\hat{\bm n},\varphi)\sin\phi _{\bm k}\right)
\tilde{f}^{{\rm EF}(0)}_{\parallel}, 
\label{eq:fEF} \\
f^{{\rm OF}(0)}_{\mu} = R_{\mu z}(\hat{\bm n},\varphi) \tilde{f}^{{\rm OF}(0)}_z .
\label{eq:fOF}
\end{gather}
Hence, the additional discrete symmetry in Eq.~\eqref{eq:pi4} arising from the ${\rm SO}(2)_{S_z+L_z}$ symmetry in Eq.~\eqref{eq:pi3} imposes a strong constraint on the possible symmetry of Cooper pair amplitudes $f^{(0)}_{\mu}$ in superfluid $^3$He-B. In particular, the OTE pairing ${\bm f}^{{\rm OF}(0)}$ that is responsible for $\chi^{\rm EP}$ is forced by the discrete symmetry to point to the surface normal direction ${\bm f}^{{\rm OF}(0)} \parallel \hat{\bm z}$. In Sec.~\ref{sec:sc}, this conclusion will be extended to pair amplitudes in superconducting states with mirror reflection symmetries.

\subsection{Ising spin anisotropy}
\label{sec:ising}

Substituting Eqs.~\eqref{eq:fEF} and \eqref{eq:fOF} into Eqs.~\eqref{eq:chiOP} and \eqref{eq:chiEP}, the spin susceptibility in Eq.~\eqref{eq:chi2} is recast into the following form:
\beq
\chi = \chi _{\rm N} + \sqrt{1-\hat{\ell}^2_{z}}\tilde{\chi}^{\rm OP}
+ \hat{\ell}_z \tilde{\chi}^{\rm EP}.
\label{eq:chi_final}
\eeq
The contributions from odd-parity and even-parity pair amplitudes are given as
\begin{gather}
\frac{\tilde{\chi}^{\rm OP}}{\chi _{\rm N}} =
-\frac{1}{\mu _{\rm n}H}{\rm Re}\left\langle 
\frac{\cos(\phi _{\bm \ell}-\phi _{\bm k})f^{{\rm OF}(1)}_0\tilde{f}^{{\rm EF}(0)\ast}_{\parallel}}{g^{(0)}_0}
\right\rangle _{\hat{\bm k},n} ,
\label{eq:chiOP2} \\
\frac{\tilde{\chi}^{\rm EP}}{\chi _{\rm N}} =
\frac{1}{\mu _{\rm n}H}{\rm Re}\left\langle 
\frac{f^{{\rm EF}(1)}_0\tilde{f}^{{\rm OF}(0)\ast}_{z}}{g^{(0)}_0}
\right\rangle _{\hat{\bm k},n}.
\label{eq:chiEP2} 
\end{gather}
Here, we introduce the unit vector, $\hat{\ell}_{\mu}(\hat{\bm n},\varphi)$~\cite{mizushimaPRL2012,mizushimaPRB2012,mizushimaJPCM2014,volovikJETP2010}
\beq
\hat{\ell}_{\mu}(\hat{\bm n},\varphi) \equiv \hat{h}_{\nu}R_{\nu\mu}(\hat{\bm n},\varphi),
\eeq
where $\phi _{\bm \ell} = \tan^{-1}(\hat{\ell}_y/\hat{\ell}_x)$ is the azimuthal angle of $\hat{\bm \ell}$.

The OTE Cooper pair at zero fields, $\tilde{f}^{{\rm OF}(0)}_{z}$, is equivalent to the low-energy surface density of states within $|E|\lesssim \Delta _0$,~\cite{higashitaniPRB2012,tsutsumiJPSJ2012}
\begin{align}
&-\frac{1}{\pi}{\rm Im}g^{(0)}_0(\hat{\bm k},z;\omega _n \rightarrow -iE+0_+)\nn \\ 
&\approx|{\rm Re}{\bm f}^{{\rm OF}(0)}(\hat{\bm k},z;\omega _n \rightarrow -iE+0_+)|,
\label{eq:equality}
\end{align}
which is always induced by the translational symmetry breaking at the surface. The surface density of states has poles at the energy
\beq
E^{(0)}_{\rm surf}({\bm k}) = \frac{\Delta _0}{k_{\rm F}}k_{\parallel},
\eeq
that is the dispersion of the surface bound states. In accordance with Eq.~\eqref{eq:equality}, the contributions from the surface bound states are contained by $\tilde{\chi}^{\rm OP}$. Although $\tilde{\chi}^{\rm OP}$ is finite, the coupling of the OTE pairing with the applied magnetic field at the surface is parameterized by $\hat{\ell}_z(\hat{\bm n},\varphi)$.

Equation \eqref{eq:chi_final} is one of the main results in this paper. This indicates that only the OTE pairs contribute to the surface spin susceptibility when $\hat{\ell}_z=0$, while $\chi$ for $\hat{\ell}_z=1$ is composed of only the ETO Cooper pairs,
\beq
\chi = \left\{
\begin{array}{ll}
\displaystyle{\chi _{\rm N}+ \tilde{\chi}^{\rm OP}} & \mbox{for $\hat{\ell}_z = 0$} \\
\\
\displaystyle{\chi _{\rm N}+ \tilde{\chi}^{\rm EP}} & \mbox{for $\hat{\ell}_z = 1$} 
\end{array}
\right. .
\eeq
In the case of bulk superfluid $^3$He-B, since the OTE pairing is absent, the spin susceptibility is given as $\chi = \chi _{\rm N}+\chi^{\rm OP}$, where $\chi^{\rm OP}<0$ suppreses the spin susceptibility. In contrast, the spin susceptibility contributed from the OTE pairs, $\chi^{\rm EP}$, is expected to increase the spin susceptibility, which comes up to $\chi > \chi _{\rm N}$.~\cite{higashitani2014} As we will discuss below, there is the critical magnetic field beyond which $\hat{\ell}_z$ becomes nonzero and the OTE pair contribute to the surface spin susceptibility.

For $^3$He-B in a slab geometry, the $\hat{\bm n}$-texture and the angle $\varphi$ are determined by the applied magnetic field, the dipole-dipole interaction arising from the magnetic moment of nuclei, and surface boundary condition. This indicates that $\hat{\ell}_z$ depends on an applied magnetic field. Let us suppose $\hat{\bm n}=\hat{\bm z}$ that is favored by the dipole-dipole interaction and specular surface boundary condition in a slab geometry. Then, one finds $\hat{\ell}_z (\hat{\bm n}= \hat{\bm z},\varphi) = \cos\theta _{\bm H}$ for a magnetic field ${\bm H}\cdot \hat{\bm z} = H \cos\theta _{\bm H}$. This configuration of the $\hat{\bm n}$-texture gives rise to the Ising anisotropy of the spin susceptibility, 
\beq
\chi = \chi _{\rm N} + \tilde{\chi}^{\rm OP}\sin\theta _{\bm H}
+ \tilde{\chi}^{\rm EP}\cos\theta _{\bm H}.
\label{eq:chiH}
\eeq
This indicates that for a magnetic field parallel to the surface ($\theta _{\bm H}=\pi/2$), although the OTE pairings exists at the surfaces, it does not couple to the applied field. The resultant spin susceptibility is contributed from only the ETO pairing, which stays about the same as that in the bulk. The OTE pairing contributes to the surface spin susceptibility when the applied field is tilted from the surface normal direction or $\hat{\ell}_z$ is nonzero. 

The $z$-component of the unit vector, $\hat{\ell}_z(\hat{\bm n},\varphi)$, has the another physical meaning that it is associated with non-trivial topological superfluidity of the B-phase under a magnetic field.~\cite{mizushimaPRL2012,mizushimaJPCM2014}
%In addition, we here consider the situation that a magnetic field is applied along $\hat{\bm h}\equiv {\bm H}/H$. Then, the quasiclassical equation contains the Zeeman term,
Since the magnetic field term in Eq.~\eqref{eq:v2} explicitly breaks the rotational symmetry in spin space as well as the time-reversal symmetry, the symmetry group of the normal $^3$He in a restricted geometry under a magnetic field is reduced to $G_{{\rm slab},H} = {\rm SO}(2)^{(\hat{\bm h})}_{\bm S} \times {\rm SO}(2)_{L_z} \times {\rm U}(1)_{\phi} \times {\rm C}$. Then, the BdG Hamiltonian of the B-phase is no longer invariant under the $\pi$-rotational symmetry, $\underline{\mathcal{U}}(\pi)\mathcal{H}({\bm k})\underline{\mathcal{U}}^{\dag}(\pi) \neq \underline{\mathcal{H}}(-\underline{\bm k})$. For $\hat{\ell}_z = 0$, however, the B-phase still holds the hidden ${\bm Z}_2$ symmetry, $H_{{\rm slab},H} = {\bm Z}_2 \times {\rm C}$. The BdG Hamiltonian is invariant under the discrete transformation given by the combination of the time-conversion operator $\underline{\mathcal{T}}$ and the $\pi$-rotation $\underline{\mathcal{U}}(\pi)$,
\beq
\underline{\mathcal{T}}~\underline{\mathcal{U}}(\pi)\underline{\mathcal{H}}({\bm k}) [\underline{\mathcal{T}}~\underline{\mathcal{U}}(\pi)]^{-1} = \underline{\mathcal{H}}(\underline{\bm k}).
\eeq
Combining the particle-hole symmetry in Eq.~\eqref{eq:phs} with the hidden ${\bm Z}_2$ symmetry, 
the chiral symmetry is preserved in the momentum space along the $k_z$-axis. As a consequence, $\hat{\ell}_z = 0$ ensures the nontrivial one-dimensional winding number, 
\beq
w = -\frac{1}{4\pi i} \int d{k}_z {\rm tr}
\left[\underline{\Gamma}\;\underline{\mathcal {H}}^{-1}({\bm k})
\partial _{k_z}\underline{\mathcal{H}}({\bm k})
\right]_{{\bm k}_{\parallel}}=2, 
\eeq
where the chiral operator is defined as $\underline{\Gamma} = \underline{\mathcal{C}}\;\underline{\mathcal{T}}\;\underline{\mathcal{U}}(\pi)$. According to the bulk-edge correspondence proven in Ref.~\onlinecite{satoPRB2011}, the winding number $w$ gives the number of the zero energy states. The chiral symmetry and nonzero winding number are responsible for the Ising anisotropy of surface spins,~\cite{mizushimaPRL2012,mizushimaJPCM2014} which implies that the surface spin operator ${\bm S}_{\rm surf}$ always points to the $\hat{\bm z}$-direction, ${\bm S}_{\rm surf} = (0,0,S^z_{\rm surf})$. The Ising spin nature of surface spins is contained in the spin susceptibility \eqref{eq:chiH} obtained from the perturbative analysis.

When $\hat{\ell}_z$ becomes nonzero, however, the ${\bm Z}_2$ symmetry is no longer held and thus the B-phase undergoes a phase transition to the non-topological phase, where the hidden ${\bm Z}_2$ symmetry is spontaneously broken by the nonzero $\hat{\ell}_z$.~\cite{mizushimaPRL2012,mizushimaJPCM2014} The nonzero $\hat{\ell}_z$ also destroys the Ising character of surface spins and the magnetic response becomes isotropic.

\subsection{Ginzburg-Landau regime}

To capture the essential part of the relation between the surface spin susceptibility and emergent Cooper pairs, we here explicitly solve the quasiclassical Eilenberger equation (\ref{eq:eilen}) within the Ginzburg-Landau approximation. In the Ginzburg-Landau regime near $T_{\rm c0}$, we may replace the diagonal component of the quasiclassical operator $g$ to the normal-state propagator $g_{\rm N} = -i\pi {\rm sgn}(\omega _n)$. For simplicity, the pair potential is assumed to be spatially uniform. In addition, we formally expand the anomalous propagator $f$ and the ${\bm d}$-vector ${\bm d}$ in powers of the applied field: $f = f^{(0)}+f^{(1)}+\cdots$ and ${\bm d} = {\bm d}^{(0)} + {\bm d}^{(1)}+\cdots$. We first solve the equation with $H=0$ and then the finite field corrections are obtained, order by order of $(\mu _n H/\Delta _0)$. In the zero field, it is obvious that the spin-singlet pair amplitudes are absent, that is, $f^{(0)}_{0}=0$. We also note that the spatially uniform pair potential is distorted by order $(\mu _{\rm n}H/\Delta _0)^2$ and we neglect ${\bm d}^{(1)}$. The pair potential at zero fields preserves the ${\rm SO}(2)_{L_z+S_z}$ symmetry, which is given in Eq.~(\ref{eq:dvec_slab}). The Cooper pair amplitudes at zero field are obtained by solving the equation for $\tilde{f}^{(0)}_{\parallel}$,~\cite{higashitaniJPSJ2014}
\beq
v_{\rm F}\hat{k}_z\partial _z \tilde{f}^{(0)}_{\parallel}
= - 2\omega _n \tilde{f}^{(0)}_{\parallel} - 2\pi {\rm sgn}(\omega _n)\Delta _{\parallel}, 
\eeq
and for $\tilde{f}^{(0)}_{z}$, 
\beq
v_{\rm F}\hat{k}_z\partial _z \tilde{f}^{(0)}_{\perp}
= - 2\omega _n \tilde{f}^{(0)}_{\perp} - 2\pi {\rm sgn}(\omega _n)\Delta _{\perp}.
\eeq
Using the specular boundary conditions at $z=0$ and $z=D$, one obtains the ETO pair amplitudes at zero field as
\begin{gather}
\tilde{f}^{{\rm EF}(0)}_{\parallel}(\theta _{\bm k},z;\omega _n) = -\pi \frac{\Delta _{\parallel}}{|\omega _n|} \sin\theta _{\bm k}, \\
\tilde{f}^{{\rm EF}(0)}_{\perp}(\theta _{\bm k},z;\omega _n) = -\pi \frac{\Delta _{\perp}\hat{k}_z}{|\omega _n|}
\left[
1 - \frac{\cosh[(z-D/2)/\lambda]}{\cosh(D/2\lambda)} 
\right],
\end{gather}
where we have introduced $\lambda = v_{\rm F}|\cos\theta _{\bm k}|/2|\omega _n|$. The OTE component emerges in the surface region as 
\begin{align}
\tilde{f}^{{\rm OF}(0)}_{\perp}(\theta _{\bm k},z;\omega _n) = -\pi \frac{\Delta _{\perp}|\hat{k}_z|}{\omega _n}
\frac{\sinh[(z-D/2)/\lambda]}{\cosh(D/2\lambda)} ,
\end{align}
and $f^{{\rm OF}(0)}_{\parallel} = 0$. 

The ESE and OSO pair amplitudes are induced by the linear Zeeman corrections. The field-induced spin-singlet pair amplitudes are governed by the following equation that are obtained from Eq.~(\ref{eq:eilen}),
\beq
iv_{\rm F}\hat{k}_z \partial _z f^{(1)}_0 = -2i\omega _n f^{(1)}_0 - \tilde{\omega}_{\rm L}\hat{\ell}_z \tilde{f}^{(0)}_{\perp}.
\eeq
The magnetic Zeeman term is parameterized by the topological order $\hat{\ell}_z$ and the effective Lamor frequency $\tilde{\omega}_{\rm L}$ is defined as 
$\tilde{\omega}_{\rm L} = \frac{2\mu _{\rm n}H}{1+F^{\rm a}_0}$. Solving the equation shown above, one finds that the ESE Cooper pair amplitude is induced at the surface $z=0$ by the magnetic Zeeman field as
\begin{align}
f^{{\rm EF}(1)}_0(\hat{\bm k},0;\omega _n) =& i\frac{\pi}{2} \hat{\ell}_z\frac{\tilde{\omega}_{\rm L}\Delta _{\perp}|\hat{k}_z|}{|\omega _n|^2} \nn \\
& \times \left[
\tanh\!\left( \frac{D}{2\lambda}\right) + \frac{D}{2\lambda}{\rm sech}^2\!\left( \frac{D}{2\lambda}\right)
\right],
\end{align}
while the OSO pair amplitude does not appear at the surface, $f^{{\rm OF}(1)}_0(\hat{\bm k},0;\omega _n) = 0$. It is also found that the intensity of the ESE Cooper pair amplitude in the central region of the system ($z\approx D/2$) exponentially decreases with increasing $D/\lambda$. Therefore, the ESE pair amplitude that are induced by the linear Zeeman corrections is localized in the surface region. 

For $D\gg \lambda$, the OTE and ESE pair amplitudes emergent at the surface are simplified as 
\begin{align}
\tilde{f}^{{\rm OF}(0)}_{\perp}(\theta _{\bm k},0;\omega _n) = \pi \frac{\Delta _{\perp}|\hat{k}_z|}{\omega _n},
\end{align}
and 
\beq
f^{{\rm EF}(1)}_0(\hat{\bm k},0;\omega _n) 
= i\frac{\pi}{2} \hat{\ell}_z|\hat{k}_z|\frac{\tilde{\omega}_{\rm L}\Delta _{\perp}}{|\omega _n|^2}.
\eeq
Substituting these expressions of OTE and ETO pair amplitudes into Eq.~(\ref{eq:chiEP2}), one obtains the first order correction to the even-parity Cooper pair contribution as 
\beq
\chi^{(1)\rm EP}_{\rm surf} = \frac{7\zeta(3)}{12(1+F^{\rm a}_0)}\left( \frac{\Delta _{\perp}}{\pi T}\right)^2   > 0,
\eeq
where $\zeta(3)$ is the Riemann zeta function. 
This clearly shows that the even-parity Cooper pairs carry the paramagnetic response $\chi^{(1)\rm EP}_{\rm surf} > 0$. Note that the odd-parity Cooper pair contribution $\chi^{(1)\rm OP}_{\rm surf}$ is absent in the Ginzburg-Landau regime. To this end, the surface spin susceptibility in the superfluid $^3$He-B is anomalously enhanced by the coupling of emergent OTE Cooper pairs to the field-induced ESE pair as 
\beq
\chi _{\rm surf} = \chi _{\rm N} + \hat{\ell}^2_z(\hat{\bm n},\varphi)\chi^{(1)\rm EP}_{\rm surf}.
\eeq
This implies that although the OTE pair amplitudes always exist in the surface of ETO superconductors and superfluids and yield paramagnetic response, they do not necessarily couple to the applied magnetic field. The topological order $\hat{\ell}_z$ that is associated with the spontaneous breaking of the hidden ${\bm Z}_2$ symmetry determines the contribution of odd-parity Cooper pairs to the surface spin susceptibility. 

%-------------------- 
\subsection{Numerical results}
\label{sec:numerical}

In the previous subsection, it has been clarified that the surface spin susceptibility is parameterized by $\hat{\ell}_z (\hat{\bm n},\varphi)$ as shown in Eq.~\eqref{eq:chi_final}. The quantity $\hat{\ell}_z$ quantifies the coupling of the OTE pairing with the applied field, leading to the Ising spin susceptibility and the zero value ensures that the B-phase stays in the symmetry protected topological phase with a gapless Majorana cone. However, the value of $\hat{\ell}_z(\hat{\bm n},\varphi)$ in equilibrium is determined by minimizing the thermodynamic potential.

We here numerically evaluate the surface spin susceptibility with self-consistent solutions. For this purpose, we solve the closed set of self-consistent equations, composed of the quasiclassical equation (\ref{eq:eilen}) for the quasiclassical propagators $\underline{g} (\hat{\bm k},{\bm r};\omega _n)$, and the equations for quasiclassical self-energies $\underline{\nu}(\hat{\bm k},{\bm r})$ and the pair potential $\Delta (\hat{\bm k},{\bm r})$. The pair potential is determined by the following gap equation with the quasiclassical propagators,
\begin{align}
d_{\mu\nu}({\bm r}) 
= & 3|g|\left\langle \hat{k}_{\nu}f_{\mu} \right\rangle _{\hat{\bm k},n}
-\tilde{g}_{\rm D}
\left( 1 + 3\delta _{\mu\nu} \right)
\left\langle \hat{k}_{\nu}{f}_{\mu} \right\rangle _{\hat{\bm k},n} \nn \\
& - 3\tilde{g}_{\rm D}\left[
\left\langle \hat{k}_{\mu}{f}_{\nu} \right\rangle _{\hat{\bm k},n}
- \left\langle \hat{k}_{\nu}{f}_{\mu} \right\rangle _{\hat{\bm k},n}
\right].
\label{eq:gapv3}
\end{align}
The pair interaction consists of the isotropic $p$-wave interaction channel with the coupling constant $g$ and the anisotropic part originating from the dipole-dipole interaction between $^3$He nuclei. The dipole interaction, which reduces the ${\rm SO}(3)_{\bm S}\!\times\!{\rm SO}(3)_{\bm L}$ symmetry to ${\rm SO}(3)_{{\bm L}+{\bm S}}$, plays a crucial role on the topological phase transition induced by a parallel magnetic field.~\cite{mizushimaPRL2012} The details on the derivation of the gap equation are described in Appendix. 

The quasiclassical self-energies $\nu _0$ and $\nu _{\mu}$ are associated with the quasiclassical Green's functions $\widehat{g}$ as
\beq
\nu _{j} (\hat{\bm k},{\bm r}) = 
\left\langle A^{(j)}(\hat{\bm k}, \hat{\bm k}^{\prime}) g_{j}(\hat{\bm k},{\bm r};i\omega _n)
\right\rangle _{\hat{\bm k}^{\prime}}, 
\label{eq:nu}
\eeq
where $A^{(j)}(\hat{\bm k}, \hat{\bm k}^{\prime})$ is expanded in terms of the Legendre polynomials $P_{\ell}$ as $A^{(j)}(\hat{\bm k}, \hat{\bm k}^{\prime})\!=\! \sum _{\ell} A^{(j)}_{\ell}P_{\ell}(\hat{\bm k}\cdot\hat{\bm k}^{\prime})$. The coefficients $A^{(j\!=\! 0)}\!=\! A^{\rm s}$ and $A^{(j\neq\! 0)} \!=\! A^{\rm a}$ are the symmetric and antisymmetric quasiparticle scattering amplitudes, which are parametrized with the Landau's Fermi liquid parameters, $F^{\rm s,a}_{\ell}$, through $F^{\rm s,a}_{\ell} \!=\! A^{\rm s,a}_{\ell}/[1-A^{\rm s,a}_{\ell}/(2\ell + 1)]$, where $F^{\rm s}_{0}\!=\! 9.3$, $F^{\rm a}_0 \!=\! 5.39$, $F^{\rm s}_1 \!=\! -0.695$, and $F^{\rm a}_1 \!=\! -0.5$. The numerical scheme to solve the self-consisten equations with a specular boundary condition in Eq.~\eqref{eq:bc} is described in Refs.~\onlinecite{mizushimaPRB2012} and \onlinecite{vorontsovPRB2003}. We fix the thickness to be $D=20\xi _0$. 

%---------------------------------------------------------
\begin{figure}[t!]
\includegraphics[width=80mm]{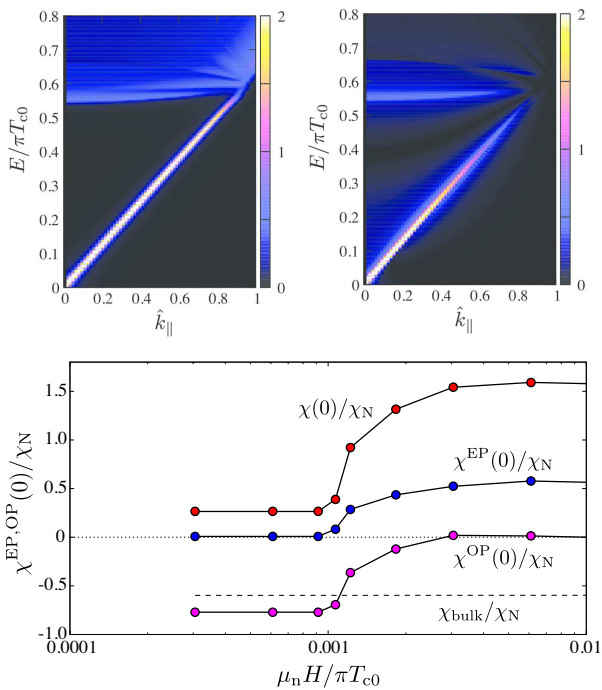}
\caption{(color online) Momentum resolved surface density of states $\mathcal{N}(\hat{\bm k},z=0;E)$ (a) and the OTE pair amplitude $|{\rm Im}f^{\rm OF}_z(\hat{\bm k},z=0;E)|$ (b). (c) Field-dependence of the surface spin susceptibilities, $\chi (0)$, $\chi^{\rm OP}(0)$, and $\chi^{\rm EP}(0)$ at $T=0.2T_{\rm c0}$. We find $\hat{\ell}_z = 0$ for $H<H_{\rm c}$, while $\hat{\ell}_z \neq 0$ when $H>H_{\rm c}$. }
\label{fig:chi}
\end{figure}
%--------------------------------------------------------/

Figures \ref{fig:chi}(a) and \ref{fig:chi}(b) show the momentum resolved surface density of states 
\beq
\mathcal{N}(\hat{\bm k},z;E) = -\frac{1}{\pi}{\rm Im}g_0(\hat{\bm k},z;\omega _n \rightarrow -iE+0_+),
\label{eq:dosk}
\eeq 
and the OTE pair amplitude $|{\rm Im}f^{\rm OF}_z(\hat{\bm k},z=0;E)|$ at the surface $z=0$, respectively. Here, we set $T=0.2T_{\rm c0}$ and $\mu _{\rm n}H = 0$, where $\hat{\bm \ell}_z = 0$. It is clearly seen that there exists the gapless surface bound state with the dispersion $E({\bm k}_{\parallel}) = \Delta _0 \hat{k}_{\parallel}$, which is called the Majorana cone, where $\hat{k}_{\parallel} = \sqrt{\hat{k}^2_x + \hat{k}^2_y}$ is the momentum in the surface. The momentum dependence of the OTE pairing traces $\mathcal{N}(\hat{\bm k},z;E)$, which indicates that the surface density of states is equivalent to the OTE pair amplitude, described in Eq.~\eqref{eq:equality}. We also find $f^{\rm OF} _x = f^{\rm OF}_y = 0$, which is consistent to Eq.~\eqref{eq:fOF}. 

The field-dependence of the surface spin susceptibility at $T=0.2T_{\rm c0}$ is plotted in Fig.~\ref{fig:chi}(c). It is found that the topological phase transition occurs at $H_{\rm c}=0.001\pi T_{\rm c0}/\mu _{\rm n} \sim 30 {\rm G}$ below which the dipole interaction favors $\hat{\ell}_z = 0$ and the ${\bm Z}_2$ symmetry is preserved even in the presence of the magnetic field. This is the topological phase with $w=2$ protected by the ${\bm Z}_2$ symmetry. At the critical field $H= H_{\rm c}$, since $\hat{\ell}_z$ becomes nonzero, the B-phase undergoes the spontaneous breaking of the ${\bm Z}_2$ symmetry which triggers the topological phase transition without closing the bulk energy gap. It turns out from Fig.~\ref{fig:chi}(c) that the topological phase transition is accompanied by the anomalous enhancement of the surface spin susceptibility.

Figure~\ref{fig:chi}(c) numerically confirms the prediction obtained from the argument of the discrete symmetry in Sec.~\ref{sec:susceptibility}. The surface spin susceptibility is divided into two contributions, $\chi = 1+\chi^{\rm EP} + \chi^{\rm OP}$, where $\chi^{\rm EP}$ and $\chi^{\rm OP}$ are the contribution from the even-parity and odd-parity pair amplitude, respectively. The behavior of the contribution from the ETO pairing, $\chi^{\rm OP}$, is understandable with the orientation of the ${\bm d}$-vector at the surface, because $\hat{\bm h}\cdot{\bm f}^{{\rm EF}(0)}$ contained in $\chi ^{\rm OP}$ indicates that the ${\bm d}$-vector at the surface is parallel to the applied field for $\hat{\ell}_z = 0$ and ${\bm d}\cdot{\bm d} = 0$ for $\hat{\ell}_z$. Therefore, the contribution of the ETO pairing reduces the spin susceptibility, compared with that in the normal $^3$He and the resultant value of $\chi^{\rm OP}$ is expected from the Yosida function, $\chi =1 + \chi^{\rm OP} \approx \chi_{\rm bulk}$. However, as the $\hat{\ell}_z$ deviates from zero (i.e., $H>H_{\rm c}$), the ETO pair amplitude tends to yield $\hat{\bm h}\cdot{\bm f}^{\rm EF} = 0$, corresponding to the situation that the ${\bm d}$-vector at the surface is normal to the applied field. Thus as $\hat{\ell}_z $ approaches $\hat{\ell}_z =1$, the spin susceptibility from the ETO pairing becomes zero ($\chi^{\rm OP}\rightarrow 0$). 

It is seen from Fig.~\ref{fig:chi}(b) that there exist the OTE pairings in the symmetry protected topological phase with $\hat{\ell}_z = 0$ (i.e., $H<H_{\rm c}$). Although the OTE pairings are responsible for the anomalous proximity effect,~\cite{higashitaniPRB2012,higashitaniPRL2013} they can not be coupled to the applied field, $\hat{\bm h}\cdot{\bm f}^{\rm OF} = 0$ as shown in Eq.~\eqref{eq:fOF}. In the non-topological phase with $\hat{\ell}_z \neq 0$, however, the OTE pairing contributes to the spin susceptibility.  

\begin{figure}[t!]
\includegraphics[width=75mm]{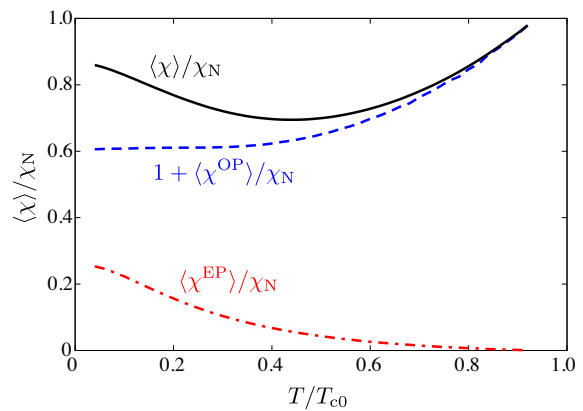}
\caption{(color online) Temperature-dependence of the spatially averaged spin susceptibilities, $\langle\chi \rangle$, $\langle\chi^{\rm OP}\rangle$, and $\langle\chi^{\rm EP}\rangle$ at $\mu _{\rm n} H = 0.009 \pi T_{\rm c0}$ and $D=20\xi _0$.}
\label{fig:chiT}
\end{figure}

As seen in Fig.~\ref{fig:chi}(c), in contrast to $\chi^{\rm OP}$, the OTE pairings ($\chi^{\rm EP}$) originating from the low-lying surface states enhance the spin susceptibility and the resultant spin susceptibility at the surface exceeds $\chi _{\rm N}$. We illustrate in Fig.~\ref{fig:chiT} the temperature dependence of the spatially averaged spin susceptibility, $\langle \chi \rangle \equiv \frac{1}{D}\int^{D}_0 \chi (z) dz$, at $\mu _{\rm n}H = 0.009\pi T_{\rm c0}$ corresponding to the non-topological phase. As discussed in Ref.~\onlinecite{mizushimaPRB2012}, the $T$-dependence of $\langle \chi \rangle $ in $^3$He-B confined to a slab exhibits the non-monotonic behavior where there exists a critical temperature below which $\langle \chi \rangle$ increases as $T$ decreases. We now identify that the increase of $\langle \chi\rangle$ in the low temperature regime of the non-topological phase reflects the coupling of the OTE pairing with the applied field, which in the topological phase with $H<H_{\rm c}$ $\langle \chi\rangle$ monotonically decreases as $T$ decreases. Hence, the anomalous behavior of the spatially averaged spin susceptibility is understandable with the concept of the odd-frequency even parity pairing, which may be observed in NMR experiments.

%-------------------- 
\section{Topological crystalline superconductors}
\label{sec:sc}

In Secs.~\ref{sec:spin} and \ref{sec:3He-B}, we have developed the theory on the relation between odd-frequency pairing and anomalous magnetic response of time-reversal invariant superfluids. In this section, we now extend this theory to time-reversal invariant superconductors that preserves the mirror reflection symmetry. We first summarize the consequence of the mirror reflection symmetry that imposes constraint on the Cooper pair amplitudes emergent in the surface.

\subsection{Mirror reflection symmetry and Ising spin anisotropy}
\label{sec:mirror}

\begin{figure}[t!]
\includegraphics[width=70mm]{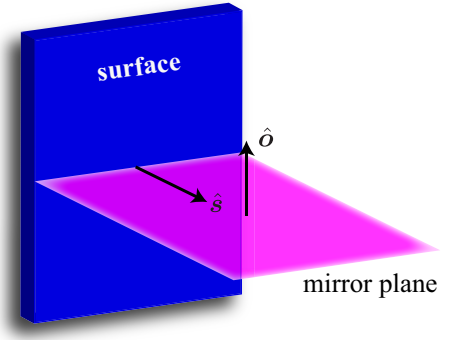}
\caption{(color online) Configuration of the specular surface and mirror reflection plane. The unit vectors, $\hat{\bm o}$ and $\hat{\bm s}$, are normal to the mirror plane and surface, respectively.}
\label{fig:mirror}
\end{figure}

Let us first suppose a topological crystalline superconductor that retains the mirror symmetry. The pair potential is even or odd under mirror reflection,
\beq
M \Delta ({\bm k},{\bm r}) M^{\rm T} = \eta \Delta (\underline{\hat{\bm k}}_{\rm M},\underline{\bm r}_{\rm M}), 
\hspace{3mm} \eta = \pm .
\label{eq:mirror2}
\eeq
The mirror reflection operator $M$ and the mirror reflected momentum $\underline{\hat{\bm k}}_{\rm M}$ have been introduced in Sec.~\ref{sec:discrete}. We here consider the configuration of the specular surface and mirror reflection plane as displayed in Fig.~\ref{fig:mirror}, where the unit vectors, $\hat{\bm o}$ and $\hat{\bm s}$, are normal to the mirror plane and surface, respectively, and we set $\hat{\bm o} \perp \hat{\bm s}$. The distance from the specular surface is denoted by $r_{\rm s} \equiv {\bm r}\cdot\hat{\bm s}$. 

%--------------------------------------------------------
%--------------------------------------------------------/

The mirror symmetry in Eq.~\eqref{eq:mirror2} topologically protects the zero energy states that are bound to the surface of TRI spin-triplet superconductors. For a superconducting state that retains the mirror symmetry \eqref{eq:mirror2}, the BdG Hamiltonian $\underline{H}({\bm k})$ satisfies the discrete symmetry in Eq.~\eqref{eq:Hmirror}. Combining the mirror symmetry $\underline{\mathcal{M}}^{\eta}$ with the time-reversal symmetry $\underline{\mathcal{T}}$ and particle-hole symmetry $\underline{\mathcal{C}}$, we have the mirror chiral symmetry~\cite{mizushimaNJP2013,tsutsumiJPSJ2013,mizushimaJPCM2014}
\beq
\{ \underline{\Gamma}, \underline{\mathcal{H}}({\bm k}_0) \} = 0, \hspace{3mm}
\underline{\Gamma} = \underline{\mathcal{C}}~\underline{\mathcal{T}}~\underline{\mathcal{M}}^{\eta},
\label{eq:chiralM}
\eeq
where ${\bm k}_0$ is defined as 
\beq
{\bm k}_0 \cdot \hat{\bm o} = 0.
\eeq
The mirror chiral symmetry enables us to define the one-dimensional winding number, 
\beq
w(k_m) = - \frac{1}{4\pi i} \int^{\pi}_{-\pi} dk_s {\rm Tr}\left[
\underline{\Gamma}~\underline{\mathcal{H}}^{-1}({\bm k}_0) \partial _{k_s} 
\underline{\mathcal{H}}({\bm k}_0)
\right],
\label{eq:wM}
\eeq
where we set $k_s \equiv {\bm k}\cdot\hat{\bm s}$ and 
\beq
k_m \equiv {\bm k}\cdot\hat{\bm m}, \hspace{3mm}
\hat{\bm m} = \hat{\bm s}\times \hat{\bm o}.
\eeq
The nontrivial value of the winding number ensures the existence of topologically protected zero energy states for the momentum ${\bm k}\parallel \hat{\bm m}$. As clarified in Sec.~\ref{sec:3He-B} and Refs.~\onlinecite{dainoPRB2012,asanoPRB2013,higashitaniPRB2012,tsutsumiJPSJ2012,stanev,hui}, the topologically protected zero modes are identical to odd-frequency Cooper pair amplitudes. 

We now derive the relation between $f(\omega _n)$ and $f(-\omega _n)$ at the surface $r_{\rm s}=0$ from the mirror symmetry that imposes the relation \eqref{eq:gmirror} on the quasiclassical propagator. In the case of the $^3$He-B, the relation is obtained in Eq.~\eqref{eq:pi4} from the $\pi$-rotation in the spin space. The spin rotation symmetry may be absent in the case of superconducting states, while the mirror symmetry arising from the crystalline symmetry can be preserved. Using the mirror symmetry \eqref{eq:gmirror} and boundary condition \eqref{eq:bc} with Eq.\eqref{eq:sym2}, one obtains the relation 
\beq
f(\hat{\bm k},r_{\rm s}=0;-\omega _n)
= - \eta M f(-\underline{\hat{\bm k}}_{\rm Ms},r_s = 0;\omega _n) M^{{\rm T}}.
\eeq
We have introduced the momentum scattered by the surface and mirror plane $\underline{\hat{\bm k}}_{\rm Ms}=\underline{\hat{\bm k}}_{\rm M} - 2 \hat{\bm s}(\hat{\bm s}\cdot\underline{\hat{\bm k}}_{\rm M})$. The constraint imposed by the mirror symmetry is then recast into 
\begin{align}
{\bm f}(-\underline{\hat{\bm k}}_{\rm Ms},0;-\omega _n)
=& \eta \left[
{\bm f}(\hat{\bm k},0;\omega _n) \right. \nn \\
& \left. - 2\hat{\bm o}\left( \hat{\bm o}\cdot{\bm f}(\hat{\bm k},0;\omega _n)\right)
\right].
\label{eq:fmirror}
\end{align}
This restricts the component of ETO and OTE pair amplitudes emergent in the surface of TRI spin-triplet superconductors. 

Let us now focus on a particular segment in the momentum space, $\hat{\bm k} \parallel \hat{\bm m}$ and $|k_m|<k_{\rm F}$, in which the topologically protected Fermi arc and odd-frequency pair amplitudes exist. This momentum segment is invariant under the mirror reflection and scattering at the surface,  $\underline{\hat{\bm k}}_{\rm Ms}=\hat{\bm k}$. Using the relation Eq.~\eqref{eq:fmirror} and focusing on the momentum segment $\hat{\bm k}=\hat{k}_m\hat{\bm m}$, one obtains the explicit form of ETO pair amplitudes for the particular region of momentum as
\beq
f^{\rm EF}_{\mu} = \frac{1}{2}(1+\eta)f_{\mu} - \eta \hat{o}_{\mu}\left( \hat{\bm o}\cdot{\bm f}\right),
\eeq
where we set $f^{\rm EF}_{\mu}\equiv f^{\rm EF}_{\mu}(\hat{\bm k}_0,0;\omega _n)$ and $f_{\mu}\equiv f_{\mu}(\hat{\bm k}_0,0;\omega _n)$. 
Similarly, the OTE pair amplitude is obtained from Eq.~\eqref{eq:fmirror} as
\beq
f^{\rm OF}_{\mu} = \frac{1}{2}(1-\eta)f_{\mu} + \eta \hat{o}_{\mu}\left( \hat{\bm o}\cdot{\bm f}\right).
\eeq
The ETO and OTE pairs emergent on the surface exhibit strong anisotropy, since the projection of ${\bm f}^{\rm EF}$ and ${\bm f}^{\rm OF}$ onto the mirror normal axis $\hat{\bm o}$ is determined by the parity of $\Delta (\hat{\bm k},{\bm r})$ under the mirror reflection, $\eta$: 
\begin{gather}
{\bm f}^{\rm EF}\cdot\hat{\bm o} = \frac{1}{2}(1-\eta) \hat{\bm o}\cdot{\bm f}, 
\label{eq:fEF_mirror} \\
{\bm f}^{\rm OF}\cdot\hat{\bm o} = \frac{1}{2}(1+\eta) \hat{\bm o}\cdot{\bm f}.
\label{eq:fOF_mirror}
\end{gather}
This implies that for $\eta = +$ ($\eta = -$), the OTE pair amplitude is forced by the mirror reflection symmetry to be parallel (perpendicular) to the mirror normal axis, $\hat{\bm o}\parallel{\bm f}^{\rm OF}$ ($\hat{\bm o}\perp{\bm f}^{\rm OF}$). 

We illustrate that the anisotropy of the emergent ETO and OTE pairs is responsible for the anisotropic magnetic response on the surface. The generic form of the surface spin susceptibility in Eq.~\eqref{eq:chi2} with Eqs.~\eqref{eq:chiOP} and \eqref{eq:chiEP} is rewritten to 
\beq
\chi \approx \chi _{\rm N} + \frac{1}{2}(1-\eta) \chi^{\rm OP} + \frac{1}{2}(1+\eta) \chi^{\rm EP},
\label{eq:chimirror}
\eeq
where the applied field is parallel to the mirror normal axis, $\hat{\bm h}\parallel\hat{\bm o}$.
We here neglect the contributions from the momentum space of $\hat{\bm k}\perp\hat{\bm m}$. In the case of $\hat{\bm h}\perp\hat{\bm o}$ where the applied magnetic field lies in the mirror plane, the surface spin susceptibility is recast to 
\beq
\chi \approx \chi _{\rm N} + \frac{1}{2}(1+\eta) \chi^{\rm OP} + \frac{1}{2}(1-\eta) \chi^{\rm EP}.
\label{eq:chimirror2}
\eeq
As discussed in Sec.~\ref{sec:3He-B}, the contribution from ETO pairs, $\chi^{\rm EP}$, are always negative and suppresses the spin susceptibility relative to $\chi _{\rm N}$. This behavior is understandable with the orientation of the ${\bm d}$-vector to the applied field. In contrast, $\chi^{\rm EP}$ is associated with the OTE pairs and anomalously enhances the surface spin susceptibility from that of the normal state. Equation~\eqref{eq:chimirror} and \eqref{eq:chimirror2} indicate that the contributions of ETO and OTE pairs to the spin susceptibility are determined by the parity of the mirror reflection symmetry, $\eta$, introduced in Eq.~\eqref{eq:mirror2}. Hence, it is from Eqs.~\eqref{eq:chimirror} and \eqref{eq:chimirror2} that a TRI spin-triplet superconductor may exhibit anomalous magnetic response, when a mirror symmetry is preserved and the OTE pair amplitudes emerge on the surface.

\subsection{Application to the $E_{1u}$ scenario of UPt$_3$}

\begin{figure}[t!]
\includegraphics[width=70mm]{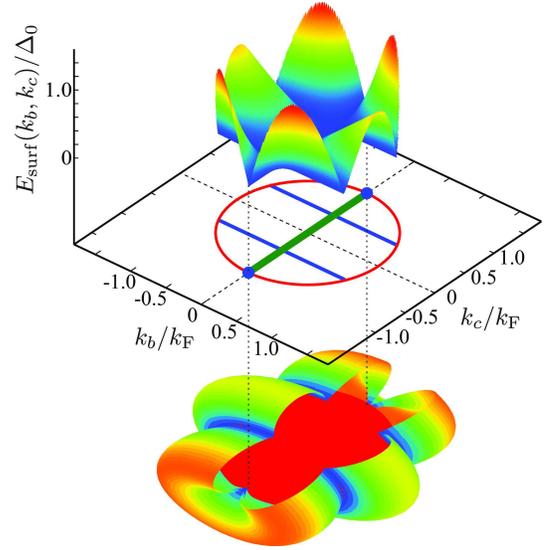}
\caption{(color online) A stereographic view of the gap function (bottom) of the $E_{1u}$ state and the dispersion (top) of the surface bound state, the Majorana valley. The topologically protected Fermi arc connects to two point nodes as shown in the central pannel.}
\label{fig:Esurf}
\end{figure}

Let us now consider the $E_{1u}$ scenario of the heavy-fermion superconductor UPt$_3$ as a prototype of topological crystalline superconductors.~\cite{machidaPRL2012,tsutsumiJPSJ2012v2,tsutsumiJPSJ2013} In the $E_{1u}$ state, the orbital part of the pair function in the B-phase that appears in the low temperature and pressure region is isotropic in the $a$-$b$ plane. The ${\bm d}$-vector is given by the ${\bm d}$-vector as~\cite{machidaPRL2012,tsutsumiJPSJ2012v2}
\beq
{\bm d}(\hat{\bm k},{\bm r}) =  \Delta _1 ({\bm r}) \lambda _a \hat{\bm b}
+ \Delta _2 ({\bm r}) \lambda _b \hat{\bm a},
\label{eq:d1}
\eeq
for the lower field $H < H_{\rm rot}$ and 
\beq
{\bm d}(\hat{\bm k},{\bm r}) =  \Delta _1 ({\bm r}) \lambda _a \hat{\bm b}
+ \Delta _2 ({\bm r}) \lambda _b \hat{\bm c},
\label{eq:d2}
\eeq
for the higher field regime $H>H_{\rm rot}$ of the B phase, where we introduce $\lambda _{a,b} = \hat{k}_{a,b} (5\hat{k}^2_c-1)$ with $\hat{k}_a \equiv \hat{\bm k}\cdot\hat{\bm a}$, $\hat{k}_b \equiv \hat{\bm k}\cdot\hat{\bm b}$, and $\hat{k}_c \equiv \hat{\bm k}\cdot\hat{\bm c}$. The gap function on three-dimensional Fermi sphere is displayed in Fig.~\ref{fig:Esurf}. The $E_{1u}$ scenario was proposed to understand the rotation of the ${\bm d}$-vectors in the Knight shift measurement for ${\bm H}\parallel {\bm c}$,~\cite{touPRL1996,touPRL1998} and is in good agreement with the recent measurement of the thermal conductivity that observes the spontaneous breaking of two-fold rotational symmetry in the C phase. Another candidate of the order parameters of UPt$_3$ has been proposed to be ${\bm d}(\hat{\bm k}) = \hat{\bm c}(\hat{k}_a\pm i \hat{k}_b)^2\hat{k}_c$ in the B phase.~\cite{joyntRMP2002,saulsAP1994} Since this pairing state spontaneously breaks the time-reversal symmetry, the present argument based on the time-reversal symmetry and the crystalline symmetry is not applicable and the Ising spin anisotropy of the surface bound states is absent. 

In the $E_{1u}$ state, the configuration of both the ${\bm d}$-vectors in Eqs.~\eqref{eq:d1} and \eqref{eq:d2} holds the mirror reflection symmetry with respect to the $a$-$c$ plane,
\beq
M \Delta (k_a,k_b,k_c) M^{\rm T} = \Delta (k_a,-k_b,k_c),
\eeq
which corresponds to the case of $\eta = +$ in Eq.~\eqref{eq:mirror2}. As shown in Fig.~\ref{fig:Esurf}, the point nodes lie on the mirror plane. The mirror operator is defined as $M=i\sigma _b$. This situation corresponds to $\hat{\bm o}=\hat{\bm b}$ and we set a specular surface to be normal to $\hat{\bm s}=\hat{\bm a}$. The mirror symmetric Hamiltonian of the UPt$_3$-B, 
\beq
\underline{\mathcal{M}}^{\eta} \underline{\mathcal{H}}({\bm k}) \underline{\mathcal{M}}^{\eta\dag}
= \underline{\mathcal{H}}(k_a,-k_b,k_c),
\eeq
holds the chiral symmetry in Eq.~\eqref{eq:chiralM}. Using this Hamiltonian, the one-dimensional winding number \eqref{eq:wM} is evaluated as
\beq
w(k_m) = \left\{
\begin{array}{ll}
2 & \hspace{3mm} \mbox{for ${\bm k}\parallel\hat{\bm c}$ and $|k_c|<k_{\rm F}$} \\
\\
0 & \hspace{3mm} \mbox{otherwise}
\end{array}
\right. .
\label{eq:wkm}
\eeq
The bulk-edge correspondence ensures the existence of the zero-energy Majorana valley two point nodes, i.e., $E_{\rm surf}({\bm k})=0$ for $|k_c| < k_{\rm F}$ and ${\bm k}\parallel\hat{\bm c}$. Therefore, as shown in Fig.~\ref{fig:Esurf}, the topologically protected Fermi arc appears in the surface of the $E_{1u}$ state of the heavy-fermion superconductor UPt$_3$-B. 

As a generic consequence of the mirror chiral symmetry, the topologically protected Fermi arc is responsible for the Ising anisotropic magnetic response that the surface bound states are gapped only by a magnetic field along the $k_b$-axis ($k_b\equiv{\bm k}\cdot\hat{\bm b}$).~\cite{mizushimaPRL2012,tsutsumiJPSJ2013,mizushimaJPCM2014} This is understandable with the emergence of OTE pairing on the surface as discussed in Sec.~\ref{sec:mirror}. It is obvious from Eq.~\eqref{eq:fOF_mirror} that for $\hat{\bm o}=\hat{\bm b}$ and $\eta = +1$, the OTE pairing emergent on the surface is restricted by the mirror reflection symmetry as
\beq
{\bm f}^{\rm OF} = (0,f^{\rm OF}_b,0),
\label{eq:fOFUPt3}
\eeq
when a magnetic field is absent. Similarly, the ETO pairing normal to the mirror reflection plane vanishes on the surface, ${\bm f}^{\rm EF}\cdot\hat{\bm b} = 0$. The anisotropy of the emergent ETO and OTE pairings is responsible for the anisotropic spin susceptibility. The surface spin susceptibility contributed from the topologically protected Fermi arc is obtained from Eq.~\eqref{eq:chimirror} as 
\beq
\chi \approx \chi _{\rm N} + \chi^{\rm EP} > \chi _{\rm N},
\eeq
for $\hat{\bm h}\parallel\hat{\bm b}$. The anomalous enhancement of the surface spin susceptibility is attributed to the contribution of odd-frequency Cooper pairs. In contrast, the surface spin susceptibility is suppressed 
\beq
\chi \approx \chi _{\rm N} + \chi^{\rm OP} < \chi _{\rm N},
\eeq
as long as the applied field lie in the mirror plane ($\hat{\bm h}\perp\hat{\bm b}$). The OTE pairing emergent on the surface is not coupled to the applied field, since ${\bm f}^{\rm OF}\cdot\hat{\bm h} = 0$ and only the ETO pairing is responsible for the magnetic response. Since one of the ${\bm d}$-vector is parallel to the applied field, the surface spin susceptibility is rather suppressed from $\chi _{\rm N}$. Hence, the topologically protected Fermi arc in the $E_{1u}$ state of the UPt$_3$-B possesses the Ising-like anisotropic magnetic response. 

Finally, to confirm the generic argument on emergent odd-frequency and Ising spin susceptibility, we here explicitly solve the quasiclassical equation \eqref{eq:eilen} for the $\hat{\bm d} = \lambda _a\hat{\bm b} + \lambda _b \hat{\bm a}$ state. Since the equation is block-diagonalized to the spin-up and down sectors, the quasiclassical propagator can be reduced in the Nambu space to
\beq
\underline{g}^{(0)} = g\underline{\tau}_z + i \left( f_1\tau_x - \sigma _z\tau _y f_2\right).
\eeq
Here, for simplicity, we neglect the Fermi liquid correction term, $\underline{\nu} = 0$.  
The quasiclassical equation for $f_1$, $f_2$, and $g$ is then obtained as
\beq
\frac{1}{2}{\bm v}_{\rm F}\cdot{\nabla}\left(
\begin{array}{c}
f_1 \\ 
f_2 \\
g
\end{array}
\right) 
= 
\left( 
\begin{array}{ccc}
0 & - \varepsilon & \mp \Delta _2 \\
\varepsilon & 0 & \Delta _1 \\
- \Delta _2 & \Delta _1 & 0 
\end{array}
\right)
\left(
\begin{array}{c}
f_1 \\ 
f_2 \\
g
\end{array}
\right).
\eeq
The homogeneous differential equation with constant coefficients can be solved for a semi-infinite system with the specular boundary condition,
\beq
\underline{g}^{(0)}(\hat{\bm k},a=0;\varepsilon) = \underline{g}^{(0)}(\underline{\hat{\bm k}},a=0;\varepsilon)
\eeq
by using the procedure developed in Refs.~\onlinecite{saulsPRB2011,haoPRB2013}.

For the $\hat{\bm d} = \lambda _a\hat{\bm b} + \lambda _b \hat{\bm a}$ state, the quasiclassical propagator $g_0(\hat{\bm k},a;\varepsilon)$ for $\varepsilon \in \mathbb{C}$ is obtained as 
\begin{align}
g^{(0)}_0(\hat{\bm k},a;\varepsilon) =& -\frac{\pi\varepsilon}{\lambda(\hat{\bm k},\varepsilon)}
\bigg[
1- \frac{\Delta^2_1(\hat{\bm k})}{\varepsilon^2-\Delta^2_2(\hat{\bm k})}
e^{-2\lambda(\hat{\bm k},\varepsilon)a/v_a}
\bigg],
\label{eq:g01}
\end{align}
where we set $v_a = v_{\rm F}\cos(\alpha)$ with $\alpha \in [-\pi/2,\pi/2]$. 
We have here introduced 
\beq
\lambda (\hat{\bm k},\varepsilon) \equiv \sqrt{|{\bm d} (\hat{\bm k})|^2 - \varepsilon^2},
\eeq
where $|{\bm d}(\hat{\bm k})| = {\bm d}(\hat{\bm k},a=\infty)$ denotes the excitation gap in the bulk. The poles of the retarded propagator $g^{R}_0(E) = g_0(\varepsilon \rightarrow E+i0_+)$ appear on the real axis 
\beq
E_{\rm surf}({\bm k}) = \pm \Delta _0 \left|5\hat{k}^2_c-1\right| \hat{k}_b 
= \pm \frac{\Delta _0}{k^3_{\rm F}} \left|5{k}^2_c-k^2_{\rm F}\right| {k}_b.
\eeq 
This is the dispersion of the surface bound states, which is displayed in Fig.~\ref{fig:Esurf}. The zero energy flat band appears for $k_b = 0$ and $|k_c| < k_{\rm F}$ as expected from the winding number in Eq.~\eqref{eq:wkm}.

%--------------------------------------------------------
%--------------------------------------------------------/

The quasiclassical propagator still has non-zero components, 
\begin{align}
g^{(0)}_c(\hat{\bm k},a;\varepsilon) = - \pi \frac{\Delta _1(\hat{\bm k})\Delta _2(\hat{\bm k})}{\epsilon^2-\Delta^2_2(\hat{\bm k})}e^{-2\lambda(\hat{\bm k},\varepsilon)a/v_a}.
\end{align}
This component of the propagator responsible for the spin current flow in the equilibrium is localized in the surface for the low-energy states $|\varepsilon| < |{\bm d}(\hat{\bm k})|$ but extended to the bulk for $|\varepsilon| > |{\bm d}(\hat{\bm k})|$.
The anomalous propagator, ${\bm f} = (f_a,f_b,f_c)$, is given as
\begin{align}
f^{(0)}_a(\hat{\bm k},a;\varepsilon) 
=& \frac{\pi\Delta _2 (\hat{\bm k})}{\lambda(\hat{\bm k},\varepsilon)}
-\frac{\pi\Delta^2_1 (\hat{\bm k})\Delta _2 (\hat{\bm k})}{\lambda(\hat{\bm k},\varepsilon)} \nn \\
& \times 
\frac{\Delta^2_1(\hat{\bm k})-\varepsilon^2}{\lambda^2\varepsilon^2-\Delta^2_1(\hat{\bm k})\Delta^2_2(\hat{\bm k})}e^{-2\lambda(\hat{\bm k},\varepsilon)a/v_a},
\end{align}
\begin{align}
f^{(0)}_b(\hat{\bm k},a;\varepsilon) 
=& \frac{\pi\Delta _1 (\hat{\bm k})}{\lambda(\hat{\bm k},\varepsilon)}
\left( 1 - e^{-2\lambda(\hat{\bm k},\varepsilon)a/v_a} \right) \nn \\
& + i\pi 
\frac{\varepsilon\Delta _1 (\hat{\bm k})(\Delta^2_1(\hat{\bm k}) - \varepsilon^2)}{\lambda^2\varepsilon^2-\Delta^2_1(\hat{\bm k})\Delta^2_2(\hat{\bm k})} 
e^{-2\lambda(\hat{\bm k},\varepsilon)a/v_a},
\end{align}
and $f^{(0)}_c(\hat{\bm k},a;\varepsilon) = 0$. Hence, at the surface $a=0$, one finds
\begin{gather}
f^{(0)}_a (\hat{\bm k},a=0;\omega _n) = f^{{\rm EF}(0)}_a(\hat{\bm k},a=0;\omega _n), \\
f^{(0)}_b (\hat{\bm k},a=0;\omega _n) = f^{{\rm OF}(0)}_b(\hat{\bm k},a=0;\omega _n),
\end{gather}
and $f^{{\rm OF}(0)}_a(\hat{\bm k},a=0;\omega _n)\!=\!f^{{\rm EF}(0)}_b(\hat{\bm k},a=0;\omega _n)\!=\! 0$. This is consistent with Eq.~\eqref{eq:fOFUPt3} that is obtained from the generic argument based on the mirror symmetry.

The emergent OTE pair amplitude is identical to the zero energy density of states on the surface. The momentum resolved density of states defined in Eq.~(\ref{eq:dosk}) is obtained from Eq.~(\ref{eq:g01}) as
\begin{align}
\mathcal{N}(\hat{\bm k},x;E) 
=& \frac{\pi}{2}\frac{\Delta^2_1(\hat{\bm k})}{\lambda(E)}
\bigg[
\delta\left(E-E_{\rm surf}(\hat{\bm k})\right) \nn \\
&+\delta\left(E+E_{\rm surf}(\hat{\bm k})\right)
\bigg]e^{-2\lambda(E)a/v_a},
\end{align}
for the bound states $|E|< |{\bm d}(\hat{\bm k})|$.
The OTE pair amplitude at $E=0$ is equivalent for the momentum resolved density of states, 
\begin{align}
\mathcal{N}(\hat{\bm k},a;E=0) 
=  \frac{1}{\pi}\left| 
{\rm Re}{\bm f}^{\rm OF}(\hat{\bm k},a;E=0)
\right| 
\end{align}
where $N(\hat{\bm k},E) = \sqrt{|{\bm d}(\hat{\bm k})|^2-E^2}$. Hence, the zero energy states protected by the mirror symmetry are equivalent to the OTE pair amplitude. The zero energy density of states on the surface is responsible for the Ising anisotropy of the surface spin susceptibility.

%\subsubsection{The $\hat{\bm d} = \lambda _a\hat{\bm b} + \lambda _b \hat{\bm c}$ state}

%-------------------- 
\section{Conclusions}
\label{sec:conclusion}

In this paper, we have examined the role of an order-two discrete symmetry on emergent Cooper pair amplitudes and magnetic response in time-reversal invariant spin-triplet superfluids and superconductors. We have first derived in Eq.~\eqref{eq:chi2} the general form of spin susceptibility within the quasiclassical formalism, which is composed of the contributions from the odd-parity pairing and even-parity pairing, $\chi^{\rm OP}$ and $\chi^{\rm EP}$. The former is associated with ETO pair amplitudes that exist in the bulk region, while the latter is the contribution from the odd-frequency Cooper pair amplitudes that emerge in the surface region as Andreev bound states. The odd-parity contribution $\chi^{\rm OP}\le 0$ is understandable with the orientation of the ${\bm d}$-vector to the applied field. In contrast, the coupling of even-parity Cooper pair amplitudes to the applied filed gives rise to the anomalous enhancement of the spin susceptibility, $\chi^{\rm EP}$. From generic argument based on the symmetry and topology of the Bogoliubov-de Gennes Hamiltonian, we have clarified that an order-two discrete symmetry preserved by the Hamiltonian imposes a strong constraint on the spin state of emergent Cooper pair amplitudes at the surface, resulting in the Ising-like anisotropy of surface spin susceptibility. 

As promising examples of time-reversal invariant topological superfluids and superconductors with an order-two discrete symmetry, we have focused on the superfluid $^3$He-B in Sec.~\ref{sec:3He-B} and heavy-fermion superconductor UPt$_3$ in Sec.~\ref{sec:sc}. The discrete symmetry in the former (latter) system originates from the $\pi$-rotation of spin and orbital spaces (mirror reflection symmetry). We have illustrated in Sec.~\ref{sec:ising} that in the case of $^3$He-B, the spin state of the emergent Cooper pair amplitudes is associated with the topological order $\hat{\ell}_z$ that characterizes the topological superfluidity of the $^3$He-B. In the symmetry protected topological phase with $\hat{\ell}_z=0$, the emergent OTE pairing is not coupled to the applied field by the order-two discrete symmetry and the surface spin susceptibility results in $\chi = \chi _{\rm N}+ \chi^{\rm OP} < \chi _{\rm N}$. For $\hat{\ell}_z>0$, however, the OTE pairing is forced to couple to the applied field, which is responsible for a large paramagnetic response as $\chi = \chi _{\rm N} + \chi^{\rm EP} > \chi _{\rm N}$. By numerically solving the quasiclassical equations, it is shown that there is the critical field beyond which the surface spin susceptibility is anomalously enhanced. This anomalous behavior is attributed to the OTE pairing enforced by the order-two discrete symmetry. The anomalous spin susceptibility is detectable with the NMR measurement in $^3$He-B confined in a slab geometry under a parallel magnetic field.~\cite{mizushimaPRB2012}

We have also illustrated that owing to the mirror reflection symmetry, the $E_{1u}$ state~\cite{machidaPRL2012,tsutsumiJPSJ2012v2,tsutsumiJPSJ2013} of the UPt$_3$-B yields the Ising-like anisotropy of the surface spin susceptibility. The anisotropic magnetic response protected by the mirror symmetry is not observed in the $E_{2u}$ state that is another possible scenario for the UPt$_3$-B.~\cite{joyntRMP2002,saulsAP1994} Since the $E_{2u}$ state spontaneously breaks time-reversal symmetry, in the case of the $E_{1u}$ state, the tunneling conductance might be sensitive to the tilting angle of the magnetic field from the mirror reflection plane. This theory is applicable to the $A_{1u}$ and $E_u$ states of Cu$_x$Bi$_2$Se$_3$ that are the three-dimensional topological superconducting state preserving the mirror reflection symmetry.~\cite{fuPRL2010,sasakiPRL2011} In this paper, however, we do not take account of the magnetic field effect coupled to the orbital motion of electrons. Full understanding of the tunneling conductance in the UPt$_3$-B still remains as a future problem.

\section*{ACKNOWLEDGMENTS}

We gratefully thank S. Higashitani, M. Sato, and Y. Tanaka for fruitful discussions and comments. This work was supported by JSPS (Nos.~25800199 and 25287085) and ``Topological Quantum Phenomena'' (No.~22103005) KAKENHI on innovation areas from MEXT.

%This work was supported by JSPS
%(No.~2074023303, 2134010303 and 22540383) and the MEXT KAKENHI (No.~22103002 and
%No.~22103005).  
%``Topological Quantum Phenomena'' (No.~22103005) KAKENHI on innovation areas from MEXT.
% 2134010303: Machida 
%No.~25800199: Mizushima (Wakate-B)
%No.~25287085: Sato (Kiban-B)

\appendix

\section{Gap equation}

We here derive the gap equation \eqref{eq:gapv3} for superfluid $^3$He, where the pairing interaction is contributed from a $p$-wave interaction and magnetic dipole-dipole interaction between $^3$He nuclei. We start with the gap equation in the Nambu-Gor'kov formalism,
\beq
\Delta _{ab}({\bm k},{\bm r}) 
= T\sum _n \int \frac{d{\bm k}}{(2\pi)^3} \mathcal{V}^{cd}_{ab}({\bm k}, {\bm k}^{\prime})\mathcal{F}_{cd}({\bm k}^{\prime},{\bm r};\omega _n),
\label{eq:gap1}
\eeq
where the repeated Roman indices imply the sum over the spins, $a,b,c,d \!=\! \uparrow,\downarrow$. The anomalous Green's functions $\mathcal{F}_{ab}$ is defined from Eq.~\eqref{eq:originalG} as
\beq
\underline{G}(\hat{\bm k},{\bm r};\omega _n)  = \left( 
\begin{array}{cc}
\mathcal{G}({\bm k},{\bm r};\omega _n)  & \mathcal{F}({\bm k},{\bm r};\omega _n)  \\ 
-\bar{\mathcal{F}}({\bm k},{\bm r};\omega _n)  & -\bar{\mathcal{G}}({\bm k},{\bm r};\omega _n) 
\end{array}
\right)
\eeq 
At the low pressure limit, the pair interaction $\mathcal{V}^{\gamma\delta}_{\alpha\beta}(\hat{\bm k}, \hat{\bm k}^{\prime})$ for $^3$He atoms is described as 
\beq
\mathcal{V}^{cd}_{ab}({\bm k}, {\bm k}^{\prime})
= 3 |g| \hat{k}_{\mu}\hat{k}^{\prime}_{\mu} \delta _{ac} \delta _{bd} 
- Q_{\mu\nu}({\bm k},{\bm k}^{\prime})\sigma^{\mu}_{ac}\sigma^{\nu}_{bd}.
\label{eq:v}
\eeq
The first term is an isotropic $p$-wave interaction with ${\rm SO}(3)_{\bm S}\!\times\!{\rm SO}(3)_{\bm L}\!\times\!{\rm U}(1)$ and the second term arises from the dipole-dipole interaction between $^3$He nuclei. The function $Q_{\mu\nu}(\hat{\bm k},\hat{\bm k}^{\prime})$ is obtained from 
\beq
\mathcal{Q}_{\mu\nu}({\bm k},{\bm k}^{\prime}) 
= 4\mu^2_{\rm n} R \int \frac{\delta _{\mu\nu}-3\hat{r}_{\mu}\hat{r}_{\nu}}{r^3}
e^{-i({\bm k}-{\bm k}^{\prime})\cdot{\bm r}}d{\bm r} ,
\eeq 
where the factor $R$ includes the contributions of high energy quasiparticles.~\cite{leggettJPC1973,leggettAP1974} 
%The dipole interaction can be expressed in terms of the partial wave series ($p$-, $f$-, and higher waves). However, since the pairing interaction between $^3$He atoms is dominated by the ${\rm SO}(3)_{\bm S}\!\times\!{\rm SO}(3)_{\bm L}\!\times\!{\rm U}(1)$ channel and the dipole interaction can be regarded as a small perturbation, we take account of only the $p$-wave contribution of $Q_{\mu\nu}(\hat{\bm k},\hat{\bm k}^{\prime})$. 

%The first term in the right-hand side of Eq.~\eqref{eq:v} yields the symmetry group ${\rm SO}(3)_{\bm S}\!\times\!{\rm SO}(3)_{\bm L}$. The dipole interaction breaks the three-dimensional rotational symmetry in each space, while it preserves the joint rotational symmetry ${\rm SO}(3)_{{\bm S}+{\bm L}}$. 

Using the partial wave expantion with the $\ell$-th spherical Bessel function $j_{\ell}(z)$ and the spherical harmonic functions ${\rm Y}_{\ell,m}$, $e^{i{\bm k}\cdot{\bm r}} \!=\! 4\pi \sum _{\ell,m}i^{\ell}j_{\ell}(kr){\rm Y}^{\ast}_{\ell,m}(\hat{\bm k}){\rm Y}_{\ell,m}(\hat{\bm r})$, the anisotropic part $Q_{\mu\nu}(\hat{\bm k},\hat{\bm k}^{\prime})$ is expaned in terms of the partial wave series
\begin{align}
&\mathcal{Q}_{\mu\nu}({\bm k},{\bm k}^{\prime}) 
= (4\pi)^2\sum _{\ell,\ell^{\prime}} 
i^{\ell^{\prime}-\ell}
\sum _{m_{\ell},m_{\ell^{\prime}}} c_{\ell\ell^{\prime}}(k,k^{\prime}) \nn \\
& \times \left\langle \ell,m_{\ell} \right| \mathcal{Q}_{\mu\nu}\left| \ell^{\prime},m_{\ell^{\prime}}
\right\rangle 
{\rm Y}_{\ell,m_{\ell}}(\hat{\bm k}){\rm Y}^{\ast}_{\ell^{\prime},m_{\ell^{\prime}}}(\hat{\bm k}^{\prime}) .
\label{eq:q}
\end{align}
Here, $c_{\ell\ell^{\prime}}(k,k^{\prime})$ describes the coupling constant, 
\beq
c_{\ell\ell^{\prime}}(k,k^{\prime}) \equiv \int^{\infty}_0 \frac{1}{r}j_{\ell}(kr)j_{\ell^{\prime}}(k^{\prime}r)dr .
\eeq
The anisotropy of the $p$-wave interaction arises from $\left\langle \ell,m_{\ell} \right| \mathcal{Q}_{\mu\nu}\left| \ell^{\prime},m^{\prime}_{\ell} \right\rangle$, which is given by
\beq
&& \hspace{-10mm}
\left\langle \ell,m_{\ell} \right| \mathcal{Q}_{\mu\nu}\left|\ell^{\prime},m_{\ell^{\prime}} \right\rangle
\equiv  \int d\hat{\bm r}
{\rm Y}^{\ast}_{\ell,m_{\ell}}(\hat{\bm r}){\rm Y}_{\ell^{\prime},m_{\ell^{\prime}}}(\hat{\bm r}) \nn \\
&& \hspace{15mm} \times 
\left(\delta _{\mu\nu}-3\hat{r}_{\mu}\hat{r}_{\nu}\right).
\label{eq:q3}
\eeq

Hence, it is seen from Eqs.~(\ref{eq:q}) and (\ref{eq:q3}) that the dipole interaction may induce the higher partial waves ($\ell \!>\! 1$). However, since the pairing interaction between $^3$He atoms is dominated by the ${\rm SO}(3)_{\bm S}\!\times\!{\rm SO}(3)_{\bm L}$ symmetric $p$-wave channel and the dipole interaction is regarded as a small perturbation, we take account of only the $p$-wave contribution of the dipole interaction, 
\begin{align}
\mathcal{Q}_{\mu\nu} ({\bm k},{\bm k}^{\prime})\sigma^{\mu}_{ac}\sigma^{\nu}_{bd} 
\approx& \tilde{g}_{\rm D} k_{\mu} \bigg[ 
\sigma^{\eta}_{ac}\sigma^{\eta}_{bd} \nn \\
& - \frac{3}{2} \left( 
\sigma^{\mu}_{ac}\sigma^{\nu}_{bd}
+ \sigma^{\nu}_{ac}\sigma^{\mu}_{bd}
\right)\bigg]k^{\prime}_{\nu},
\label{eq:QQ}
\end{align}
where $\tilde{g}_{\rm D}\!\equiv\! \frac{24\pi}{5}\gamma^2R$. %For convenience, 
%\beq
%d_{\mu}(\hat{\bm k},{\bm r}) = 
%-\frac{1}{2}{\rm Tr}\left[i\sigma _y\sigma _{\mu} 
%\Delta ({\bm k},{\bm r})
%\right] \equiv d_{\mu\nu}({\bm r})\hat{k}_{\nu}
%\eeq
Substituting Eq.~\eqref{eq:QQ} into Eq.~\eqref{eq:v}, the gap equation \eqref{eq:gap1} is recast into 
\begin{align}
d_{\mu\nu}({\bm r}) 
= & -3|g|\left\langle \hat{k}_{\nu}\mathcal{F}_{\mu} \right\rangle _{\hat{\bm k},n}
+\tilde{g}_{\rm D}
\left( 1 + 3\delta _{\mu\nu} \right)
\left\langle \hat{k}_{\nu}\mathcal{F}_{\mu} \right\rangle _{\hat{\bm k},n} \nn \\
& + 3\tilde{g}_{\rm D}\left[
\left\langle \hat{k}_{\mu}\mathcal{F}_{\nu} \right\rangle _{\hat{\bm k},n}
- \left\langle \hat{k}_{\nu}\mathcal{F}_{\mu} \right\rangle _{\hat{\bm k},n}
\right],
\label{eq:gapNG}
\end{align}
where $d_{\mu}({\bm k},{\bm r}) = d_{\mu\nu}({\bm r})\hat{k}_{\nu}= - \frac{1}{2}{\rm Tr}[ i\sigma _y \sigma _{\mu} \Delta ({\bm k},{\bm r})]$.
%The factor $R$ includes the contributions of high energy quasiparticles.~\cite{leggettJPC1973,leggettAP1974} 
%The details are described in Appendix. 
In the frame of the quasiclassical theory, the gap equation (\ref{eq:gap1}) is expressed in terms of the quasiclassical propagators $f_{\mu}$ as
\begin{align}
d_{\mu\nu}({\bm r}) 
= & 3|g|\left\langle \hat{k}_{\nu}f_{\mu} \right\rangle _{\hat{\bm k},n}
-\tilde{g}_{\rm D}
\left( 1 + 3\delta _{\mu\nu} \right)
\left\langle \hat{k}_{\nu}{f}_{\mu} \right\rangle _{\hat{\bm k},n} \nn \\
& - 3\tilde{g}_{\rm D}\left[
\left\langle \hat{k}_{\mu}{f}_{\nu} \right\rangle _{\hat{\bm k},n}
- \left\langle \hat{k}_{\nu}{f}_{\mu} \right\rangle _{\hat{\bm k},n}
\right].
\label{eq:gap}
\end{align}
The dipole interaction characterized with the effective coupling constant $\tilde{g}_{\rm D}$ induces distortion to the isotropic $p$-wave interaction with the coupling constant $g$.  

%---------------------------------------------------------
\begin{figure}[t!]
\includegraphics[width=80mm]{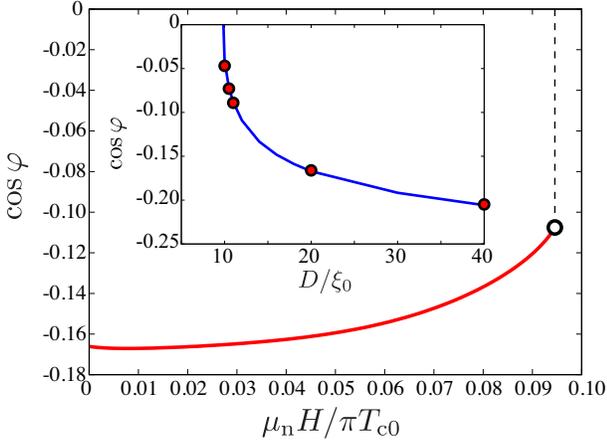}
\caption{(color online) Field-dependence of the stable Leggett angle $\varphi$ for ${\bm H}\parallel\hat{\bm z}$ at $T=0.2T_{\rm c0}$ and $D=20\xi _0$, where the $\hat{\bm n}$-vector is aligned to the $\hat{\bm z}$-axis. The first-order A-B phase transition occurs at $\mu _nH/\pi T_{\rm c0} \approx 0.09$ corresponding to $0.36$T. The inset of (a) shows the $D$-dependence of the Leggett angle $\varphi$ at $H=0$ and $T=0.2T_{\rm c0}$, where the solid curve is obtained from Eq.~(\ref{eq:varphi}). }
\label{fig:leggett}
\end{figure}
%--------------------------------------------------------/

Let us now show that for spatially uniform $d_{\mu\nu}$, the gap equation \eqref{eq:gap} reproduces the Leggett angle in the thermodynamic limit. The quasiclassical propagator $\underline{g}$ at the limit is obtained from Eq.~(\ref{eq:eilen}) with the normalization condition as 
\beq
\underline{g}(\hat{\bm k},{\bm r};\omega _n) = -\pi 
\frac{i\omega _n \underline{\tau}_0+\underline{\Delta}(\hat{\bm k})}{\sqrt{\omega^2_n+|{\bm d}(\hat{\bm k})|^2}}, 
\eeq
where for simplicity a magnetic field is assumed to be absent. Then, the gap equation (\ref{eq:gap}) is recast into 
\begin{align}
& d_{\mu\nu} \left( 1 - 3|g|J^{(2)}_{\nu}\right) \nn \\
& = - \tilde{g}_{\rm D} \left(
3\delta _{\mu\nu}d_{\gamma\gamma}J^{(2)}_{\gamma} 
-2 d_{\mu\nu}J^{(2)}_{\nu}
+ 3 d_{\nu\mu}J^{(2)}_{\mu}
\right),
\label{eq:gap2}
\end{align}
where 
\beq
J^{(n)}_{\mu}\equiv \left\langle \hat{k}^n_{\mu}\frac{\pi}{\sqrt{\omega^2_n+|{\bm d}(\hat{\bm k})|^2}}
\right\rangle _{\hat{\bm k},n}.
\eeq 
We regard the contribution of the dipole interaction as a small perturbation, which reduces the gap equation (\ref{eq:gap2}) to
\begin{align}
d_{\mu\nu} \left( 1 - 3|g|J^{(2)}_{\nu}\right) 
= - \frac{\tilde{g}_{\rm D}}{3|g|} \left(
3\delta _{\mu\nu}d_{\gamma\gamma}
-2 d_{\mu\nu}
+ 3 d_{\nu\mu}
\right),
\label{eq:gap3}
\end{align}
where the higher order terms on $\tilde{g}_{\rm D}$ are neglected. Without loss of generality, the rotation axis is set to be along the $z$-axis in the thermodynamic limit. Then, the order parameter of the B-phase distorted by the dipole interaction is described as $d_{\mu\nu} \!=\! R_{\mu\nu}(\hat{n}_z,\varphi)\Delta _{\nu}$, where $\Delta _{x} \!=\! \Delta _y \!\equiv \! \Delta _{\parallel}$ and $\Delta _z \!=\! \Delta _{\perp}$. Substituting this order parameters into Eq.~(\ref{eq:gap3}), one finds the set of three equations for the amplitudes $\Delta _{\parallel}$ and $\Delta _{\perp}$ and angle $\varphi$, 
\begin{gather}
\frac{\Delta _{\perp}}{\Delta _{\parallel}}\left( 1-3|g|J^{(2)}_z\right) 
= -\tilde{g}_{\rm D}J^{(2)}_z\left( 3\cos\!\varphi+2\frac{\Delta _{\perp}}{\Delta _{\parallel}}\right), 
\label{eq:gap4a} \\
1-\frac{3}{2}|g|(J^{(0)}-J^{(2)}_z)= 
\frac{\tilde{g}_{\rm D}}{2}(J^{(2)}_z-J^{(0)})\left( 7\cos\!\varphi + 3\frac{\Delta _{\perp}}{\Delta _{\parallel}}\right), 
\label{eq:gap4b} \\
1 - \frac{3}{2}|g|(J^{(0)}-J^{(2)}_z)
= -\frac{5}{2}\tilde{g}_{\rm D}(J^{(0)}-J^{(2)}_z).
\label{eq:gap4c}
\end{gather}
Equations (\ref{eq:gap4b}) and (\ref{eq:gap4c}) determine the relative angle $\varphi$ of the rotation matrix within the lowest order on $\tilde{g}_{\rm D}$ as
\beq
\varphi = \cos^{-1}\left(-\frac{1}{4}\frac{\Delta _{\perp}}{\Delta _{\parallel}}\right).
\eeq
This is consistent to the so-called Leggett angle that was obtained in Refs.~\onlinecite{tewordtPL1976,tewordtJLTP1979,schopohlJLTP1982,fishmanPRB1987}. The orientation of the $\hat{\bm n}$-vector is determined by the competition between the dipole interaction, magnetic field, and pair breaking effect at the surface.

We also microscopically determine the angle $\varphi$ that minimizes the thermodynamic potential with the self-consistent solutions in a slab geometry. In the Ginzburg-Landau regime, the dipole energy density $f_{\rm dip}$ is obtained as~\cite{vollhardt}
\beq
f_{\rm dip} = \frac{1}{5}\lambda _{\rm D}\mathcal{N}_{\rm F} \left(
d^{\ast}_{\mu\mu}d_{\nu\nu} + d^{\ast}_{\mu\nu}d_{\nu\mu} -\frac{2}{3}d^{\ast}_{\mu\nu}d_{\mu\nu}
\right),
\label{eq:fdip}
\eeq
where $\lambda _{\rm D}$ a dimensionless dipole coupling parameter and approximately independent of pressure. The value is estimated as $\lambda _{\rm D}\sim 5\times 10^{-7}$.~\cite{vollhardt} For simplicity, we here consider the case of a perpendicular magnetic field ${\bm H}\parallel\hat{\bm z}$ in a slab geometry. In this situation, the order parameter is given as
\begin{align}
d_{\mu\nu}(z) = R_{\mu\eta}(\hat{\bm n},\varphi) \left[
\Delta _{\parallel}(z) \left( \delta _{\eta ,\nu} - \hat{z}_{\eta}\hat{z}_{\nu}\right)
+ \Delta _{\perp}(z) \hat{z}_{\eta}\hat{z}_{\nu}
\right],
\end{align}
where $\hat{\bm z}$ is the unit vector normal to the surface. For ${\bm H}\parallel\hat{\bm z}$, the $\hat{\bm n}$-vector is always oriented to the surface normal direction, regardless of the value of $H$. Substituting this order parameter to Eq.~(\ref{eq:fdip}), one finds that the local minimum of $\mathcal{F}_{\rm dip}\equiv \int f_{\rm dip}d{z}$ exists at~\cite{mizushimaJPCM2014}
\beq
\hat{\bm n} = (0,0,1), 
\eeq
and 
\beq
\varphi = \cos^{-1}\left( -\frac{1}{4}\frac{\langle \Delta_{\parallel}(z)\Delta _{\perp}(z)\rangle}{\langle \Delta^2_{\parallel}\rangle} \right).
\label{eq:varphi}
\eeq
This solution is obtained by solving $\partial \mathcal{F}_{\rm dip}/\partial \hat{\bm n}=0$ and $\partial \mathcal{F}_{\rm dip}/\partial \varphi = 0$. We have here introduced the spatial average over the slab, $\langle \cdots\rangle = D^{-1}\int^{D}_0 \cdots dz$.

The main panel of Fig.~\ref{fig:leggett} shows the field-dependence of $\varphi$ for ${\bm H}\parallel\hat{\bm z}$ and the inset is the $D$-dependence at zero fields, where we fix $T=0.2T_{\rm c0}$ and $D=20\xi _0$. As seen in the inset of Fig.~\ref{fig:leggett}, the $D$-dependence of $\varphi$ at zero fields is in good agreement with Eq.~(\ref{eq:varphi}) in the basis of the Ginzburg-Landau analysis. The angle $\varphi$ approaches zero at the critical thickness $D\approx 9.8\xi _0$ that the A-B phase transition occurs. As seen in the main panel of Fig.~\ref{fig:leggett}, the angle $\varphi$ is relatively insensitive to the increase of the applied field $H$.

\bibliography{topology}

\begin{thebibliography}{68}
\expandafter\ifx\csname natexlab\endcsname\relax\def\natexlab#1{#1}\fi
\expandafter\ifx\csname bibnamefont\endcsname\relax
  \def\bibnamefont#1{#1}\fi
\expandafter\ifx\csname bibfnamefont\endcsname\relax
  \def\bibfnamefont#1{#1}\fi
\expandafter\ifx\csname citenamefont\endcsname\relax
  \def\citenamefont#1{#1}\fi
\expandafter\ifx\csname url\endcsname\relax
  \def\url#1{\texttt{#1}}\fi
\expandafter\ifx\csname urlprefix\endcsname\relax\def\urlprefix{URL }\fi
\providecommand{\bibinfo}[2]{#2}
\providecommand{\eprint}[2][]{\url{#2}}

\bibitem[{\citenamefont{Tanaka et~al.}(2012)\citenamefont{Tanaka, Sato, and
  Nagaosa}}]{tanakaJPSJ2012}
\bibinfo{author}{\bibfnamefont{Y.}~\bibnamefont{Tanaka}},
  \bibinfo{author}{\bibfnamefont{M.}~\bibnamefont{Sato}}, \bibnamefont{and}
  \bibinfo{author}{\bibfnamefont{N.}~\bibnamefont{Nagaosa}},
  \bibinfo{journal}{J. Phys. Soc. Jpn.} \textbf{\bibinfo{volume}{81}},
  \bibinfo{pages}{011013} (\bibinfo{year}{2012}).

\bibitem[{\citenamefont{Kashiwaya and Tanaka}(2000)}]{kashiwayaRPP2000}
\bibinfo{author}{\bibfnamefont{S.}~\bibnamefont{Kashiwaya}} \bibnamefont{and}
  \bibinfo{author}{\bibfnamefont{Y.}~\bibnamefont{Tanaka}},
  \bibinfo{journal}{Rep. Prog. Phys.} \textbf{\bibinfo{volume}{63}},
  \bibinfo{pages}{1641} (\bibinfo{year}{2000}).

\bibitem[{\citenamefont{Tanaka and Kashiwaya}(1996)}]{tanakaPRB1996}
\bibinfo{author}{\bibfnamefont{Y.}~\bibnamefont{Tanaka}} \bibnamefont{and}
  \bibinfo{author}{\bibfnamefont{S.}~\bibnamefont{Kashiwaya}},
  \bibinfo{journal}{Phys. Rev. B} \textbf{\bibinfo{volume}{53}},
  \bibinfo{pages}{R11957} (\bibinfo{year}{1996}).

\bibitem[{\citenamefont{Tanaka and Kashiwaya}(1997)}]{tanakaPRB1997}
\bibinfo{author}{\bibfnamefont{Y.}~\bibnamefont{Tanaka}} \bibnamefont{and}
  \bibinfo{author}{\bibfnamefont{S.}~\bibnamefont{Kashiwaya}},
  \bibinfo{journal}{Phys. Rev. B} \textbf{\bibinfo{volume}{56}},
  \bibinfo{pages}{892} (\bibinfo{year}{1997}).

\bibitem[{\citenamefont{Barash et~al.}(1996)\citenamefont{Barash, Burkhardt,
  and Rainer}}]{barashPRL1996}
\bibinfo{author}{\bibfnamefont{Y.~S.} \bibnamefont{Barash}},
  \bibinfo{author}{\bibfnamefont{H.}~\bibnamefont{Burkhardt}},
  \bibnamefont{and} \bibinfo{author}{\bibfnamefont{D.}~\bibnamefont{Rainer}},
  \bibinfo{journal}{Phys. Rev. Lett.} \textbf{\bibinfo{volume}{77}},
  \bibinfo{pages}{4070} (\bibinfo{year}{1996}).

\bibitem[{\citenamefont{Kashiwaya et~al.}(1999)\citenamefont{Kashiwaya, Tanaka,
  Yoshida, and Beasley}}]{kashiwayaPRB1999}
\bibinfo{author}{\bibfnamefont{S.}~\bibnamefont{Kashiwaya}},
  \bibinfo{author}{\bibfnamefont{Y.}~\bibnamefont{Tanaka}},
  \bibinfo{author}{\bibfnamefont{N.}~\bibnamefont{Yoshida}}, \bibnamefont{and}
  \bibinfo{author}{\bibfnamefont{M.~R.} \bibnamefont{Beasley}},
  \bibinfo{journal}{Phys. Rev. B} \textbf{\bibinfo{volume}{60}},
  \bibinfo{pages}{3572} (\bibinfo{year}{1999}).

\bibitem[{\citenamefont{Tanaka et~al.}(2007{\natexlab{a}})\citenamefont{Tanaka,
  Golubov, Kashiwaya, and Ueda}}]{tanakaPRL2007}
\bibinfo{author}{\bibfnamefont{Y.}~\bibnamefont{Tanaka}},
  \bibinfo{author}{\bibfnamefont{A.~A.} \bibnamefont{Golubov}},
  \bibinfo{author}{\bibfnamefont{S.}~\bibnamefont{Kashiwaya}},
  \bibnamefont{and} \bibinfo{author}{\bibfnamefont{M.}~\bibnamefont{Ueda}},
  \bibinfo{journal}{Phys. Rev. Lett.} \textbf{\bibinfo{volume}{99}},
  \bibinfo{pages}{037005} (\bibinfo{year}{2007}{\natexlab{a}}).

\bibitem[{\citenamefont{Higashitani}(1997)}]{higashitaniJPSJ1997}
\bibinfo{author}{\bibfnamefont{S.}~\bibnamefont{Higashitani}},
  \bibinfo{journal}{J. Phys. Soc. Jpn.} \textbf{\bibinfo{volume}{66}},
  \bibinfo{pages}{2556} (\bibinfo{year}{1997}).

\bibitem[{\citenamefont{Walter et~al.}(1998)\citenamefont{Walter, Prusseit,
  Semerad, Kinder, Assmann, Huber, Burkhardt, Rainer, and
  Sauls}}]{WalterPRL1998}
\bibinfo{author}{\bibfnamefont{H.}~\bibnamefont{Walter}},
  \bibinfo{author}{\bibfnamefont{W.}~\bibnamefont{Prusseit}},
  \bibinfo{author}{\bibfnamefont{R.}~\bibnamefont{Semerad}},
  \bibinfo{author}{\bibfnamefont{H.}~\bibnamefont{Kinder}},
  \bibinfo{author}{\bibfnamefont{W.}~\bibnamefont{Assmann}},
  \bibinfo{author}{\bibfnamefont{H.}~\bibnamefont{Huber}},
  \bibinfo{author}{\bibfnamefont{H.}~\bibnamefont{Burkhardt}},
  \bibinfo{author}{\bibfnamefont{D.}~\bibnamefont{Rainer}}, \bibnamefont{and}
  \bibinfo{author}{\bibfnamefont{J.~A.} \bibnamefont{Sauls}},
  \bibinfo{journal}{Phys. Rev. Lett.} \textbf{\bibinfo{volume}{80}},
  \bibinfo{pages}{3598} (\bibinfo{year}{1998}).

\bibitem[{\citenamefont{Yokoyama et~al.}(2011)\citenamefont{Yokoyama, Tanaka,
  and Nagaosa}}]{yokoyamaPRL2011}
\bibinfo{author}{\bibfnamefont{T.}~\bibnamefont{Yokoyama}},
  \bibinfo{author}{\bibfnamefont{Y.}~\bibnamefont{Tanaka}}, \bibnamefont{and}
  \bibinfo{author}{\bibfnamefont{N.}~\bibnamefont{Nagaosa}},
  \bibinfo{journal}{Phys. Rev. Lett.} \textbf{\bibinfo{volume}{106}},
  \bibinfo{pages}{246601} (\bibinfo{year}{2011}).

\bibitem[{\citenamefont{Asano et~al.}(2011)\citenamefont{Asano, Golubov,
  Fominov, and Tanaka}}]{asanoPRL2011}
\bibinfo{author}{\bibfnamefont{Y.}~\bibnamefont{Asano}},
  \bibinfo{author}{\bibfnamefont{A.~A.} \bibnamefont{Golubov}},
  \bibinfo{author}{\bibfnamefont{Y.~V.} \bibnamefont{Fominov}},
  \bibnamefont{and} \bibinfo{author}{\bibfnamefont{Y.}~\bibnamefont{Tanaka}},
  \bibinfo{journal}{Phys. Rev. Lett.} \textbf{\bibinfo{volume}{107}},
  \bibinfo{pages}{087001} (\bibinfo{year}{2011}).

\bibitem[{\citenamefont{Bergeret et~al.}(2001)\citenamefont{Bergeret, Volkov,
  and Efetov}}]{bergeretPRL2001}
\bibinfo{author}{\bibfnamefont{F.~S.} \bibnamefont{Bergeret}},
  \bibinfo{author}{\bibfnamefont{A.~F.} \bibnamefont{Volkov}},
  \bibnamefont{and} \bibinfo{author}{\bibfnamefont{K.~B.}
  \bibnamefont{Efetov}}, \bibinfo{journal}{Phys. Rev. Lett.}
  \textbf{\bibinfo{volume}{86}}, \bibinfo{pages}{4096} (\bibinfo{year}{2001}).

\bibitem[{\citenamefont{Tanaka et~al.}(2005)\citenamefont{Tanaka, Asano,
  Golubov, and Kashiwaya}}]{tanakaPRB2005}
\bibinfo{author}{\bibfnamefont{Y.}~\bibnamefont{Tanaka}},
  \bibinfo{author}{\bibfnamefont{Y.}~\bibnamefont{Asano}},
  \bibinfo{author}{\bibfnamefont{A.~A.} \bibnamefont{Golubov}},
  \bibnamefont{and}
  \bibinfo{author}{\bibfnamefont{S.}~\bibnamefont{Kashiwaya}},
  \bibinfo{journal}{Phys. Rev. B} \textbf{\bibinfo{volume}{72}},
  \bibinfo{pages}{140503} (\bibinfo{year}{2005}).

\bibitem[{\citenamefont{Tanaka and Golubov}(2007)}]{tanakaPRL2007v2}
\bibinfo{author}{\bibfnamefont{Y.}~\bibnamefont{Tanaka}} \bibnamefont{and}
  \bibinfo{author}{\bibfnamefont{A.~A.} \bibnamefont{Golubov}},
  \bibinfo{journal}{Phys. Rev. Lett.} \textbf{\bibinfo{volume}{98}},
  \bibinfo{pages}{037003} (\bibinfo{year}{2007}).

\bibitem[{\citenamefont{Linder et~al.}(2009)\citenamefont{Linder, Yokoyama,
  Sudb\o{}, and Eschrig}}]{linderPRL2009}
\bibinfo{author}{\bibfnamefont{J.}~\bibnamefont{Linder}},
  \bibinfo{author}{\bibfnamefont{T.}~\bibnamefont{Yokoyama}},
  \bibinfo{author}{\bibfnamefont{A.}~\bibnamefont{Sudb\o{}}}, \bibnamefont{and}
  \bibinfo{author}{\bibfnamefont{M.}~\bibnamefont{Eschrig}},
  \bibinfo{journal}{Phys. Rev. Lett.} \textbf{\bibinfo{volume}{102}},
  \bibinfo{pages}{107008} (\bibinfo{year}{2009}).

\bibitem[{\citenamefont{Linder et~al.}(2010)\citenamefont{Linder, Sudb\o{},
  Yokoyama, Grein, and Eschrig}}]{linderPRB2010}
\bibinfo{author}{\bibfnamefont{J.}~\bibnamefont{Linder}},
  \bibinfo{author}{\bibfnamefont{A.}~\bibnamefont{Sudb\o{}}},
  \bibinfo{author}{\bibfnamefont{T.}~\bibnamefont{Yokoyama}},
  \bibinfo{author}{\bibfnamefont{R.}~\bibnamefont{Grein}}, \bibnamefont{and}
  \bibinfo{author}{\bibfnamefont{M.}~\bibnamefont{Eschrig}},
  \bibinfo{journal}{Phys. Rev. B} \textbf{\bibinfo{volume}{81}},
  \bibinfo{pages}{214504} (\bibinfo{year}{2010}).

\bibitem[{\citenamefont{Higashitani}(2014{\natexlab{a}})}]{higashitani2014}
\bibinfo{author}{\bibfnamefont{S.}~\bibnamefont{Higashitani}},
  \bibinfo{journal}{Phys. Rev. B} \textbf{\bibinfo{volume}{89}},
  \bibinfo{pages}{184505} (\bibinfo{year}{2014}{\natexlab{a}}).

\bibitem[{asa()}]{asano2014}
\bibinfo{note}{Y. Asano, Y. V. Fominov, and Y. Tanaka, arXiv:1407.2325}.

\bibitem[{\citenamefont{Berezinskii}(1974)}]{berezinskiiJETP1974}
\bibinfo{author}{\bibfnamefont{V.~L.} \bibnamefont{Berezinskii}},
  \bibinfo{journal}{JETP Lett.} \textbf{\bibinfo{volume}{20}},
  \bibinfo{pages}{287} (\bibinfo{year}{1974}).

\bibitem[{\citenamefont{Chung and Zhang}(2009)}]{chungPRL2009}
\bibinfo{author}{\bibfnamefont{S.~B.} \bibnamefont{Chung}} \bibnamefont{and}
  \bibinfo{author}{\bibfnamefont{S.-C.} \bibnamefont{Zhang}},
  \bibinfo{journal}{Phys. Rev. Lett.} \textbf{\bibinfo{volume}{103}},
  \bibinfo{pages}{235301} (\bibinfo{year}{2009}).

\bibitem[{\citenamefont{Nagato et~al.}(2009)\citenamefont{Nagato, Higashitani,
  and Nagai}}]{nagatoJPSJ2009}
\bibinfo{author}{\bibfnamefont{Y.}~\bibnamefont{Nagato}},
  \bibinfo{author}{\bibfnamefont{S.}~\bibnamefont{Higashitani}},
  \bibnamefont{and} \bibinfo{author}{\bibfnamefont{K.}~\bibnamefont{Nagai}},
  \bibinfo{journal}{J. Phys. Soc. Jpn.} \textbf{\bibinfo{volume}{78}},
  \bibinfo{pages}{123603} (\bibinfo{year}{2009}).

\bibitem[{\citenamefont{Shindou et~al.}(2010)\citenamefont{Shindou, Furusaki,
  and Nagaosa}}]{shindouPRB2010}
\bibinfo{author}{\bibfnamefont{R.}~\bibnamefont{Shindou}},
  \bibinfo{author}{\bibfnamefont{A.}~\bibnamefont{Furusaki}}, \bibnamefont{and}
  \bibinfo{author}{\bibfnamefont{N.}~\bibnamefont{Nagaosa}},
  \bibinfo{journal}{Phys. Rev. B} \textbf{\bibinfo{volume}{82}},
  \bibinfo{pages}{180505} (\bibinfo{year}{2010}).

\bibitem[{\citenamefont{Volovik}(2010)}]{volovikJETP2010}
\bibinfo{author}{\bibfnamefont{G.}~\bibnamefont{Volovik}},
  \bibinfo{journal}{JETP Lett.} \textbf{\bibinfo{volume}{91}},
  \bibinfo{pages}{201} (\bibinfo{year}{2010}).

\bibitem[{\citenamefont{Mizushima et~al.}(2012)\citenamefont{Mizushima, Sato,
  and Machida}}]{mizushimaPRL2012}
\bibinfo{author}{\bibfnamefont{T.}~\bibnamefont{Mizushima}},
  \bibinfo{author}{\bibfnamefont{M.}~\bibnamefont{Sato}}, \bibnamefont{and}
  \bibinfo{author}{\bibfnamefont{K.}~\bibnamefont{Machida}},
  \bibinfo{journal}{Phys. Rev. Lett.} \textbf{\bibinfo{volume}{109}},
  \bibinfo{pages}{165301} (\bibinfo{year}{2012}).

\bibitem[{\citenamefont{Tsutsumi et~al.}(2013)\citenamefont{Tsutsumi, Ishikawa,
  Kawakami, Mizushima, Sato, Ichioka, and Machida}}]{tsutsumiJPSJ2013}
\bibinfo{author}{\bibfnamefont{Y.}~\bibnamefont{Tsutsumi}},
  \bibinfo{author}{\bibfnamefont{M.}~\bibnamefont{Ishikawa}},
  \bibinfo{author}{\bibfnamefont{T.}~\bibnamefont{Kawakami}},
  \bibinfo{author}{\bibfnamefont{T.}~\bibnamefont{Mizushima}},
  \bibinfo{author}{\bibfnamefont{M.}~\bibnamefont{Sato}},
  \bibinfo{author}{\bibfnamefont{M.}~\bibnamefont{Ichioka}}, \bibnamefont{and}
  \bibinfo{author}{\bibfnamefont{K.}~\bibnamefont{Machida}},
  \bibinfo{journal}{J. Phys. Soc. Jpn.} \textbf{\bibinfo{volume}{82}},
  \bibinfo{pages}{113707} (\bibinfo{year}{2013}).

\bibitem[{shi()}]{shiozaki2014}
\bibinfo{note}{K. Shiozaki and M. Sato, arXiv:1403.3331.}

\bibitem[{miz({\natexlab{a}})}]{mizushimaJPCM2014}
\bibinfo{note}{T. Mizushima, Y. Tsutsumi, M. Sato, and K. Machida, in
  preparation.}

\bibitem[{\citenamefont{Mizushima and Sato}(2013)}]{mizushimaNJP2013}
\bibinfo{author}{\bibfnamefont{T.}~\bibnamefont{Mizushima}} \bibnamefont{and}
  \bibinfo{author}{\bibfnamefont{M.}~\bibnamefont{Sato}}, \bibinfo{journal}{New
  J. Phys.} \textbf{\bibinfo{volume}{15}}, \bibinfo{pages}{075010}
  (\bibinfo{year}{2013}).

\bibitem[{\citenamefont{Sato and Fujimoto}(2009)}]{satoPRB2009}
\bibinfo{author}{\bibfnamefont{M.}~\bibnamefont{Sato}} \bibnamefont{and}
  \bibinfo{author}{\bibfnamefont{S.}~\bibnamefont{Fujimoto}},
  \bibinfo{journal}{Phys. Rev. B} \textbf{\bibinfo{volume}{79}},
  \bibinfo{pages}{094504} (\bibinfo{year}{2009}).

\bibitem[{\citenamefont{Fu and Berg}(2010)}]{fuPRL2010}
\bibinfo{author}{\bibfnamefont{L.}~\bibnamefont{Fu}} \bibnamefont{and}
  \bibinfo{author}{\bibfnamefont{E.}~\bibnamefont{Berg}},
  \bibinfo{journal}{Phys. Rev. Lett.} \textbf{\bibinfo{volume}{105}},
  \bibinfo{pages}{097001} (\bibinfo{year}{2010}).

\bibitem[{\citenamefont{Sasaki et~al.}(2011)\citenamefont{Sasaki, Kriener,
  Segawa, Yada, Tanaka, Sato, and Ando}}]{sasakiPRL2011}
\bibinfo{author}{\bibfnamefont{S.}~\bibnamefont{Sasaki}},
  \bibinfo{author}{\bibfnamefont{M.}~\bibnamefont{Kriener}},
  \bibinfo{author}{\bibfnamefont{K.}~\bibnamefont{Segawa}},
  \bibinfo{author}{\bibfnamefont{K.}~\bibnamefont{Yada}},
  \bibinfo{author}{\bibfnamefont{Y.}~\bibnamefont{Tanaka}},
  \bibinfo{author}{\bibfnamefont{M.}~\bibnamefont{Sato}}, \bibnamefont{and}
  \bibinfo{author}{\bibfnamefont{Y.}~\bibnamefont{Ando}},
  \bibinfo{journal}{Phys. Rev. Lett.} \textbf{\bibinfo{volume}{107}},
  \bibinfo{pages}{217001} (\bibinfo{year}{2011}).

\bibitem[{\citenamefont{Koren et~al.}(2011)\citenamefont{Koren, Kirzhner,
  Lahoud, Chashka, and Kanigel}}]{kirzhner1}
\bibinfo{author}{\bibfnamefont{G.}~\bibnamefont{Koren}},
  \bibinfo{author}{\bibfnamefont{T.}~\bibnamefont{Kirzhner}},
  \bibinfo{author}{\bibfnamefont{E.}~\bibnamefont{Lahoud}},
  \bibinfo{author}{\bibfnamefont{K.~B.} \bibnamefont{Chashka}},
  \bibnamefont{and} \bibinfo{author}{\bibfnamefont{A.}~\bibnamefont{Kanigel}},
  \bibinfo{journal}{Phys. Rev. B} \textbf{\bibinfo{volume}{84}},
  \bibinfo{pages}{224521} (\bibinfo{year}{2011}).

\bibitem[{\citenamefont{Kirzhner et~al.}(2012)\citenamefont{Kirzhner, Lahoud,
  Chaska, Salman, and Kanigel}}]{kirzhner2}
\bibinfo{author}{\bibfnamefont{T.}~\bibnamefont{Kirzhner}},
  \bibinfo{author}{\bibfnamefont{E.}~\bibnamefont{Lahoud}},
  \bibinfo{author}{\bibfnamefont{K.~B.} \bibnamefont{Chaska}},
  \bibinfo{author}{\bibfnamefont{Z.}~\bibnamefont{Salman}}, \bibnamefont{and}
  \bibinfo{author}{\bibfnamefont{A.}~\bibnamefont{Kanigel}},
  \bibinfo{journal}{Phys. Rev. B} \textbf{\bibinfo{volume}{86}},
  \bibinfo{pages}{064517} (\bibinfo{year}{2012}).

\bibitem[{\citenamefont{Koren and Kirzhner}(2012)}]{kirzhner3}
\bibinfo{author}{\bibfnamefont{G.}~\bibnamefont{Koren}} \bibnamefont{and}
  \bibinfo{author}{\bibfnamefont{T.}~\bibnamefont{Kirzhner}},
  \bibinfo{journal}{Phys. Rev. B} \textbf{\bibinfo{volume}{86}},
  \bibinfo{pages}{144508} (\bibinfo{year}{2012}).

\bibitem[{\citenamefont{Peng et~al.}(2013)\citenamefont{Peng, De, Lv, Wei, and
  Chu}}]{peng}
\bibinfo{author}{\bibfnamefont{H.}~\bibnamefont{Peng}},
  \bibinfo{author}{\bibfnamefont{D.}~\bibnamefont{De}},
  \bibinfo{author}{\bibfnamefont{B.}~\bibnamefont{Lv}},
  \bibinfo{author}{\bibfnamefont{F.}~\bibnamefont{Wei}}, \bibnamefont{and}
  \bibinfo{author}{\bibfnamefont{C.-W.} \bibnamefont{Chu}},
  \bibinfo{journal}{Phys. Rev. B} \textbf{\bibinfo{volume}{88}},
  \bibinfo{pages}{024515} (\bibinfo{year}{2013}).

\bibitem[{\citenamefont{Levy et~al.}(2013)\citenamefont{Levy, Zhang, Ha,
  Sharifi, Talin, Kuk, and Stroscio}}]{levy}
\bibinfo{author}{\bibfnamefont{N.}~\bibnamefont{Levy}},
  \bibinfo{author}{\bibfnamefont{T.}~\bibnamefont{Zhang}},
  \bibinfo{author}{\bibfnamefont{J.}~\bibnamefont{Ha}},
  \bibinfo{author}{\bibfnamefont{F.}~\bibnamefont{Sharifi}},
  \bibinfo{author}{\bibfnamefont{A.~A.} \bibnamefont{Talin}},
  \bibinfo{author}{\bibfnamefont{Y.}~\bibnamefont{Kuk}}, \bibnamefont{and}
  \bibinfo{author}{\bibfnamefont{J.~A.} \bibnamefont{Stroscio}},
  \bibinfo{journal}{Phys. Rev. Lett.} \textbf{\bibinfo{volume}{110}},
  \bibinfo{pages}{117001} (\bibinfo{year}{2013}).

\bibitem[{miz({\natexlab{b}})}]{mizushima2014}
\bibinfo{note}{T. Mizushima, A. Yamakage, M. Sato, and Y. Tanaka,
  arXiv:1311.2768.}

\bibitem[{\citenamefont{Daino et~al.}(2012)\citenamefont{Daino, Ichioka,
  Mizushima, and Tanaka}}]{dainoPRB2012}
\bibinfo{author}{\bibfnamefont{T.}~\bibnamefont{Daino}},
  \bibinfo{author}{\bibfnamefont{M.}~\bibnamefont{Ichioka}},
  \bibinfo{author}{\bibfnamefont{T.}~\bibnamefont{Mizushima}},
  \bibnamefont{and} \bibinfo{author}{\bibfnamefont{Y.}~\bibnamefont{Tanaka}},
  \bibinfo{journal}{Phys. Rev. B} \textbf{\bibinfo{volume}{86}},
  \bibinfo{pages}{064512} (\bibinfo{year}{2012}).

\bibitem[{\citenamefont{Asano and Tanaka}(2013)}]{asanoPRB2013}
\bibinfo{author}{\bibfnamefont{Y.}~\bibnamefont{Asano}} \bibnamefont{and}
  \bibinfo{author}{\bibfnamefont{Y.}~\bibnamefont{Tanaka}},
  \bibinfo{journal}{Phys. Rev. B} \textbf{\bibinfo{volume}{87}},
  \bibinfo{pages}{104513} (\bibinfo{year}{2013}).

\bibitem[{\citenamefont{Higashitani et~al.}(2012)\citenamefont{Higashitani,
  Matsuo, Nagato, Nagai, Murakawa, Nomura, and Okuda}}]{higashitaniPRB2012}
\bibinfo{author}{\bibfnamefont{S.}~\bibnamefont{Higashitani}},
  \bibinfo{author}{\bibfnamefont{S.}~\bibnamefont{Matsuo}},
  \bibinfo{author}{\bibfnamefont{Y.}~\bibnamefont{Nagato}},
  \bibinfo{author}{\bibfnamefont{K.}~\bibnamefont{Nagai}},
  \bibinfo{author}{\bibfnamefont{S.}~\bibnamefont{Murakawa}},
  \bibinfo{author}{\bibfnamefont{R.}~\bibnamefont{Nomura}}, \bibnamefont{and}
  \bibinfo{author}{\bibfnamefont{Y.}~\bibnamefont{Okuda}},
  \bibinfo{journal}{Phys. Rev. B} \textbf{\bibinfo{volume}{85}},
  \bibinfo{pages}{024524} (\bibinfo{year}{2012}).

\bibitem[{\citenamefont{Tsutsumi and Machida}(2012)}]{tsutsumiJPSJ2012}
\bibinfo{author}{\bibfnamefont{Y.}~\bibnamefont{Tsutsumi}} \bibnamefont{and}
  \bibinfo{author}{\bibfnamefont{K.}~\bibnamefont{Machida}},
  \bibinfo{journal}{J. Phys. Soc. Jpn.} \textbf{\bibinfo{volume}{81}},
  \bibinfo{pages}{074607} (\bibinfo{year}{2012}).

\bibitem[{\citenamefont{Stanev and Galitski}(2014)}]{stanev}
\bibinfo{author}{\bibfnamefont{V.}~\bibnamefont{Stanev}} \bibnamefont{and}
  \bibinfo{author}{\bibfnamefont{V.}~\bibnamefont{Galitski}},
  \bibinfo{journal}{Phys. Rev. B} \textbf{\bibinfo{volume}{89}},
  \bibinfo{pages}{174521} (\bibinfo{year}{2014}).

\bibitem[{\citenamefont{Hui et~al.}(2014)\citenamefont{Hui, Sau, and
  Das~Sarma}}]{hui}
\bibinfo{author}{\bibfnamefont{H.-Y.} \bibnamefont{Hui}},
  \bibinfo{author}{\bibfnamefont{J.~D.} \bibnamefont{Sau}}, \bibnamefont{and}
  \bibinfo{author}{\bibfnamefont{S.}~\bibnamefont{Das~Sarma}},
  \bibinfo{journal}{Phys. Rev. B} \textbf{\bibinfo{volume}{90}},
  \bibinfo{pages}{064516} (\bibinfo{year}{2014}).

\bibitem[{\citenamefont{Ueno et~al.}(2013)\citenamefont{Ueno, Yamakage, Tanaka,
  and Sato}}]{uenoPRL2013}
\bibinfo{author}{\bibfnamefont{Y.}~\bibnamefont{Ueno}},
  \bibinfo{author}{\bibfnamefont{A.}~\bibnamefont{Yamakage}},
  \bibinfo{author}{\bibfnamefont{Y.}~\bibnamefont{Tanaka}}, \bibnamefont{and}
  \bibinfo{author}{\bibfnamefont{M.}~\bibnamefont{Sato}},
  \bibinfo{journal}{Phys. Rev. Lett.} \textbf{\bibinfo{volume}{111}},
  \bibinfo{pages}{087002} (\bibinfo{year}{2013}).

\bibitem[{\citenamefont{Serene and Rainer}(1983)}]{serene}
\bibinfo{author}{\bibfnamefont{J.}~\bibnamefont{Serene}} \bibnamefont{and}
  \bibinfo{author}{\bibfnamefont{D.}~\bibnamefont{Rainer}},
  \bibinfo{journal}{Phys. Rep.} \textbf{\bibinfo{volume}{101}},
  \bibinfo{pages}{221 } (\bibinfo{year}{1983}).

\bibitem[{vol()}]{vollhardt}
\bibinfo{note}{D. Vollhardt and P. W\"{o}lfle, {\it The Superfluid Phases of
  Helium 3} (Taylor and Francis, London, 1990)}.

\bibitem[{\citenamefont{Mizushima}(2012)}]{mizushimaPRB2012}
\bibinfo{author}{\bibfnamefont{T.}~\bibnamefont{Mizushima}},
  \bibinfo{journal}{Phys. Rev. B} \textbf{\bibinfo{volume}{86}},
  \bibinfo{pages}{094518} (\bibinfo{year}{2012}).

\bibitem[{\citenamefont{Higashitani et~al.}(2013)\citenamefont{Higashitani,
  Takeuchi, Matsuo, Nagato, and Nagai}}]{higashitaniPRL2013}
\bibinfo{author}{\bibfnamefont{S.}~\bibnamefont{Higashitani}},
  \bibinfo{author}{\bibfnamefont{H.}~\bibnamefont{Takeuchi}},
  \bibinfo{author}{\bibfnamefont{S.}~\bibnamefont{Matsuo}},
  \bibinfo{author}{\bibfnamefont{Y.}~\bibnamefont{Nagato}}, \bibnamefont{and}
  \bibinfo{author}{\bibfnamefont{K.}~\bibnamefont{Nagai}},
  \bibinfo{journal}{Phys. Rev. Lett.} \textbf{\bibinfo{volume}{110}},
  \bibinfo{pages}{175301} (\bibinfo{year}{2013}).

\bibitem[{\citenamefont{Tanaka et~al.}(2007{\natexlab{b}})\citenamefont{Tanaka,
  Tanuma, and Golubov}}]{tanakaPRB2007}
\bibinfo{author}{\bibfnamefont{Y.}~\bibnamefont{Tanaka}},
  \bibinfo{author}{\bibfnamefont{Y.}~\bibnamefont{Tanuma}}, \bibnamefont{and}
  \bibinfo{author}{\bibfnamefont{A.~A.} \bibnamefont{Golubov}},
  \bibinfo{journal}{Phys. Rev. B} \textbf{\bibinfo{volume}{76}},
  \bibinfo{pages}{054522} (\bibinfo{year}{2007}{\natexlab{b}}).

\bibitem[{\citenamefont{Yokoyama et~al.}(2008)\citenamefont{Yokoyama, Tanaka,
  and Golubov}}]{yokoyamaPRB2008}
\bibinfo{author}{\bibfnamefont{T.}~\bibnamefont{Yokoyama}},
  \bibinfo{author}{\bibfnamefont{Y.}~\bibnamefont{Tanaka}}, \bibnamefont{and}
  \bibinfo{author}{\bibfnamefont{A.~A.} \bibnamefont{Golubov}},
  \bibinfo{journal}{Phys. Rev. B} \textbf{\bibinfo{volume}{78}},
  \bibinfo{pages}{012508} (\bibinfo{year}{2008}).

\bibitem[{\citenamefont{Yokoyama et~al.}(2010)\citenamefont{Yokoyama, Ichioka,
  and Tanaka}}]{yokoyamaJPSJ2010}
\bibinfo{author}{\bibfnamefont{T.}~\bibnamefont{Yokoyama}},
  \bibinfo{author}{\bibfnamefont{M.}~\bibnamefont{Ichioka}}, \bibnamefont{and}
  \bibinfo{author}{\bibfnamefont{Y.}~\bibnamefont{Tanaka}},
  \bibinfo{journal}{Journal of the Physical Society of Japan}
  \textbf{\bibinfo{volume}{79}}, \bibinfo{pages}{034702}
  (\bibinfo{year}{2010}).

\bibitem[{\citenamefont{Sato et~al.}(2011)\citenamefont{Sato, Tanaka, Yada, and
  Yokoyama}}]{satoPRB2011}
\bibinfo{author}{\bibfnamefont{M.}~\bibnamefont{Sato}},
  \bibinfo{author}{\bibfnamefont{Y.}~\bibnamefont{Tanaka}},
  \bibinfo{author}{\bibfnamefont{K.}~\bibnamefont{Yada}}, \bibnamefont{and}
  \bibinfo{author}{\bibfnamefont{T.}~\bibnamefont{Yokoyama}},
  \bibinfo{journal}{Phys. Rev. B} \textbf{\bibinfo{volume}{83}},
  \bibinfo{pages}{224511} (\bibinfo{year}{2011}).

\bibitem[{\citenamefont{Higashitani}(2014{\natexlab{b}})}]{higashitaniJPSJ2014}
\bibinfo{author}{\bibfnamefont{S.}~\bibnamefont{Higashitani}},
  \bibinfo{journal}{J. Phys. Soc. Jpn.} \textbf{\bibinfo{volume}{83}},
  \bibinfo{pages}{075002} (\bibinfo{year}{2014}{\natexlab{b}}).

\bibitem[{\citenamefont{Vorontsov and Sauls}(2003)}]{vorontsovPRB2003}
\bibinfo{author}{\bibfnamefont{A.~B.} \bibnamefont{Vorontsov}}
  \bibnamefont{and} \bibinfo{author}{\bibfnamefont{J.~A.} \bibnamefont{Sauls}},
  \bibinfo{journal}{Phys. Rev. B} \textbf{\bibinfo{volume}{68}},
  \bibinfo{pages}{064508} (\bibinfo{year}{2003}).

\bibitem[{\citenamefont{Machida et~al.}(2012)\citenamefont{Machida, Itoh, So,
  Izawa, Haga, Yamamoto, Kimura, Onuki, Tsutsumi, and
  Machida}}]{machidaPRL2012}
\bibinfo{author}{\bibfnamefont{Y.}~\bibnamefont{Machida}},
  \bibinfo{author}{\bibfnamefont{A.}~\bibnamefont{Itoh}},
  \bibinfo{author}{\bibfnamefont{Y.}~\bibnamefont{So}},
  \bibinfo{author}{\bibfnamefont{K.}~\bibnamefont{Izawa}},
  \bibinfo{author}{\bibfnamefont{Y.}~\bibnamefont{Haga}},
  \bibinfo{author}{\bibfnamefont{E.}~\bibnamefont{Yamamoto}},
  \bibinfo{author}{\bibfnamefont{N.}~\bibnamefont{Kimura}},
  \bibinfo{author}{\bibfnamefont{Y.}~\bibnamefont{Onuki}},
  \bibinfo{author}{\bibfnamefont{Y.}~\bibnamefont{Tsutsumi}}, \bibnamefont{and}
  \bibinfo{author}{\bibfnamefont{K.}~\bibnamefont{Machida}},
  \bibinfo{journal}{Phys. Rev. Lett.} \textbf{\bibinfo{volume}{108}},
  \bibinfo{pages}{157002} (\bibinfo{year}{2012}).

\bibitem[{\citenamefont{Tsutsumi et~al.}(2012)\citenamefont{Tsutsumi, Machida,
  Ohmi, and Ozaki}}]{tsutsumiJPSJ2012v2}
\bibinfo{author}{\bibfnamefont{Y.}~\bibnamefont{Tsutsumi}},
  \bibinfo{author}{\bibfnamefont{K.}~\bibnamefont{Machida}},
  \bibinfo{author}{\bibfnamefont{T.}~\bibnamefont{Ohmi}}, \bibnamefont{and}
  \bibinfo{author}{\bibfnamefont{M.-a.} \bibnamefont{Ozaki}},
  \bibinfo{journal}{J. Phys. Soc. Jpn.} \textbf{\bibinfo{volume}{81}},
  \bibinfo{pages}{074717} (\bibinfo{year}{2012}).

\bibitem[{\citenamefont{Tou et~al.}(1996)\citenamefont{Tou, Kitaoka, Asayama,
  Kimura, {\=O}nuki, Yamamoto, and Maezawa}}]{touPRL1996}
\bibinfo{author}{\bibfnamefont{H.}~\bibnamefont{Tou}},
  \bibinfo{author}{\bibfnamefont{Y.}~\bibnamefont{Kitaoka}},
  \bibinfo{author}{\bibfnamefont{K.}~\bibnamefont{Asayama}},
  \bibinfo{author}{\bibfnamefont{N.}~\bibnamefont{Kimura}},
  \bibinfo{author}{\bibfnamefont{Y.}~\bibnamefont{{\=O}nuki}},
  \bibinfo{author}{\bibfnamefont{E.}~\bibnamefont{Yamamoto}}, \bibnamefont{and}
  \bibinfo{author}{\bibfnamefont{K.}~\bibnamefont{Maezawa}},
  \bibinfo{journal}{Phys. Rev. Lett.} \textbf{\bibinfo{volume}{77}},
  \bibinfo{pages}{1374} (\bibinfo{year}{1996}).

\bibitem[{\citenamefont{Tou et~al.}(1998)\citenamefont{Tou, Kitaoka, Ishida,
  Asayama, Kimura, {\=O}nuki, Yamamoto, Haga, and Maezawa}}]{touPRL1998}
\bibinfo{author}{\bibfnamefont{H.}~\bibnamefont{Tou}},
  \bibinfo{author}{\bibfnamefont{Y.}~\bibnamefont{Kitaoka}},
  \bibinfo{author}{\bibfnamefont{K.}~\bibnamefont{Ishida}},
  \bibinfo{author}{\bibfnamefont{K.}~\bibnamefont{Asayama}},
  \bibinfo{author}{\bibfnamefont{N.}~\bibnamefont{Kimura}},
  \bibinfo{author}{\bibfnamefont{Y.}~\bibnamefont{{\=O}nuki}},
  \bibinfo{author}{\bibfnamefont{E.}~\bibnamefont{Yamamoto}},
  \bibinfo{author}{\bibfnamefont{Y.}~\bibnamefont{Haga}}, \bibnamefont{and}
  \bibinfo{author}{\bibfnamefont{K.}~\bibnamefont{Maezawa}},
  \bibinfo{journal}{Phys. Rev. Lett.} \textbf{\bibinfo{volume}{80}},
  \bibinfo{pages}{3129} (\bibinfo{year}{1998}).

\bibitem[{\citenamefont{Joynt and Taillefer}(2002)}]{joyntRMP2002}
\bibinfo{author}{\bibfnamefont{R.}~\bibnamefont{Joynt}} \bibnamefont{and}
  \bibinfo{author}{\bibfnamefont{L.}~\bibnamefont{Taillefer}},
  \bibinfo{journal}{Rev. Mod. Phys.} \textbf{\bibinfo{volume}{74}},
  \bibinfo{pages}{235} (\bibinfo{year}{2002}).

\bibitem[{\citenamefont{Sauls}(1994)}]{saulsAP1994}
\bibinfo{author}{\bibfnamefont{J.~A.} \bibnamefont{Sauls}},
  \bibinfo{journal}{Adv. Phys.} \textbf{\bibinfo{volume}{43}},
  \bibinfo{pages}{113} (\bibinfo{year}{1994}).

\bibitem[{\citenamefont{Sauls}(2011)}]{saulsPRB2011}
\bibinfo{author}{\bibfnamefont{J.~A.} \bibnamefont{Sauls}},
  \bibinfo{journal}{Phys. Rev. B} \textbf{\bibinfo{volume}{84}},
  \bibinfo{pages}{214509} (\bibinfo{year}{2011}).

\bibitem[{\citenamefont{Wu and Sauls}(2013)}]{haoPRB2013}
\bibinfo{author}{\bibfnamefont{H.}~\bibnamefont{Wu}} \bibnamefont{and}
  \bibinfo{author}{\bibfnamefont{J.~A.} \bibnamefont{Sauls}},
  \bibinfo{journal}{Phys. Rev. B} \textbf{\bibinfo{volume}{88}},
  \bibinfo{pages}{184506} (\bibinfo{year}{2013}).

\bibitem[{\citenamefont{Leggett}(1973)}]{leggettJPC1973}
\bibinfo{author}{\bibfnamefont{A.~J.} \bibnamefont{Leggett}},
  \bibinfo{journal}{J. Phys. C} \textbf{\bibinfo{volume}{6}},
  \bibinfo{pages}{3187} (\bibinfo{year}{1973}).

\bibitem[{\citenamefont{Leggett}(1974)}]{leggettAP1974}
\bibinfo{author}{\bibfnamefont{A.~J.} \bibnamefont{Leggett}},
  \bibinfo{journal}{Ann. Phys. (N.Y.)} \textbf{\bibinfo{volume}{85}},
  \bibinfo{pages}{11} (\bibinfo{year}{1974}).

\bibitem[{\citenamefont{Tewordt and Einzel}(1976)}]{tewordtPL1976}
\bibinfo{author}{\bibfnamefont{L.}~\bibnamefont{Tewordt}} \bibnamefont{and}
  \bibinfo{author}{\bibfnamefont{D.}~\bibnamefont{Einzel}},
  \bibinfo{journal}{Phys. Lett.} \textbf{\bibinfo{volume}{56A}},
  \bibinfo{pages}{97} (\bibinfo{year}{1976}).

\bibitem[{\citenamefont{Tewordt and Schopohl}(1979)}]{tewordtJLTP1979}
\bibinfo{author}{\bibfnamefont{L.}~\bibnamefont{Tewordt}} \bibnamefont{and}
  \bibinfo{author}{\bibfnamefont{N.}~\bibnamefont{Schopohl}},
  \bibinfo{journal}{J. Low Temp. Phys.} \textbf{\bibinfo{volume}{37}},
  \bibinfo{pages}{421} (\bibinfo{year}{1979}).

\bibitem[{\citenamefont{Schopohl}(1982)}]{schopohlJLTP1982}
\bibinfo{author}{\bibfnamefont{N.}~\bibnamefont{Schopohl}},
  \bibinfo{journal}{J. Low Temp. Phys.} \textbf{\bibinfo{volume}{49}},
  \bibinfo{pages}{347} (\bibinfo{year}{1982}).

\bibitem[{\citenamefont{Fishman}(1987)}]{fishmanPRB1987}
\bibinfo{author}{\bibfnamefont{R.~S.} \bibnamefont{Fishman}},
  \bibinfo{journal}{Phys. Rev. B} \textbf{\bibinfo{volume}{36}},
  \bibinfo{pages}{79} (\bibinfo{year}{1987}).

\end{thebibliography}

\end{document}